\documentclass[11pt]{article}
\usepackage{a4wide,amsmath,amssymb,comment,nicefrac,footnpag,cite}
\usepackage{times}
\def\di{{\rm d}}
\def\zi{\mathbb{Z}}

\long\def\symbolfootnote[#1]#2{\begingroup%
\def\thefootnote{\fnsymbol{footnote}}\footnote[#1]{#2}\endgroup}

\begin{document}

\begin{titlepage}

\centerline{}
\vspace{2cm}
\centerline{\LARGE \bf T-Duality in Gauged Linear Sigma-Models with Torsion}

\vskip 1.6cm
\centerline{Dan Isra\"el\symbolfootnote[1]{Email:
israel@lpthe.jussieu.fr}}
\vskip 1cm

\centerline{\sl Sorbonne Universit\'es, UPMC Univ Paris 06, UMR 7589, LPTHE, F-75005, 
Paris, France}
\centerline{\sl CNRS, UMR 7589, LPTHE, F-75005, Paris, France}

\vskip 2.4cm

\centerline{\bf Abstract} \vskip 0.5cm 
(0,2) gauged linear sigma models with torsion, corresponding to principal torus bundles over warped CY bases, 
provide a useful framework for getting exact statements about perturbative dualities in the presence of fluxes. In this context 
we first study dualities mapping the torus fiber onto itself, implying the existence of 
quantization constraints on the torus moduli for consistency. Second, we investigate dualities mixing the principal torus 
bundle with the gauge bundle, relating the torsional GLSMs to ordinary ones corresponding to CY compactifications 
with non-standard embeddings, namely two classes of models with different target-space topologies.

\noindent

\vfill

%\keywords{Heterotic strings, Flux compactifications, Gauged linear sigma models, T-duality.}
%\arxivnumber{xxx}

\end{titlepage}

%%%%%%%%%%%%%%%%%%%%%%%%%%%%%%%%%%%%%%%%%%%%%%%%%%%%%%%%%%%%%%%%%%%%%%%%%%%%%%%%%%%%%%%%%%%%%%%%%%%%%%%%%%%%%%%%%%%%%%%%%%%%%%%%%%%%%%%%%%%%%%%%%%%%%%%%%%%%%%%%%%%

\section{Introduction}

Gauged linear sigma models (GLSMs), introduced by Witten~\cite{Witten:1993yc}, are defined as $(2,2)$ or $(0,2)$ supersymmetric two-dimensional gauge theories with Abelian gauge groups, that are designed to flow in the infrared, in their geometric phase, 
to superconformal non-linear sigma-models suitable for supersymmetric string compactifications. They provide a very 
efficient tool for computing topological quantities in worldsheet theories for Calabi-Yau compactifications, 
even far away from the special loci in moduli space where a conformal field theory description is known; 
for a recent review see~\cite{McOrist:2010ae}. Recently, $(0,2)$ linear 
sigma models with torsion, describing heterotic compactifications with flux, triggered a renewal of interest for this 
subject~\cite{Adams:2006kb,Lapan:2007zz,Adams:2009av,Adams:2009zg,NibbelinkGroot:2010wm,Quigley:2011pv,Blaszczyk:2011ib,Quigley:2012gq,Adams:2012sh,Melnikov:2012nm}. 
GLSMs are even more useful in this context, since so little is known about the space of torsional compactifications.

Moving away from the phenomenologically unappealing case of CY compactifications with standard embedding of the spin connection in 
the gauge connection is possible by changing the gauge bundle or adding three-form flux threading the compact geometry. 
These two aspects are actually tied together by the modified heterotic Bianchi identity, 
which proves notoriously arduous to solve. Explicit solutions of the supersymmetry equations at order $\alpha'$ that were written 
down long time ago~\cite{Strominger:1986uh} appeared only rather recently. A class of solutions, consisting in principal torus bundles over 
a warped CY base, were first discovered using string dualities~\cite{Dasgupta:1999ss}, then, following~\cite{Goldstein:2002pg}, 
shown to be provide actual solutions of the supersymmetry equations~\cite{Fu:2006vj}. Other classes of solutions belong to a 
mostly uncharted territory. Indeed, having a non-trivial three-form flux results in the metric loosing K\"ahlerity  and 
being conformally balanced instead of Calabi-Yau~\cite{Ivanov:2000ai,LopesCardoso:2002hd}. 
Some non-compact conifold solutions with torsion were nevertheless found in~\cite{Carlevaro:2009jx} as exact worldsheet CFT backgrounds, 
and supergravities generalizations thereof were studied recently in~\cite{Chen:2013nma}.

In order to deal with compact torsional heterotic backgrounds, gauged linear sigma models are the only available models 
at present. Adding torsion to  $(0,2)$ GLSMs was first understood in~\cite{Adams:2006kb,Lapan:2007zz,Adams:2009av,Adams:2009zg}, 
where the torus bundle solutions mentioned in the last paragraph were obtained by canceling the worldsheet gauge anomaly 
coming from the non-standard gauge bundle against classically non gauge-invariant axial couplings of chiral multiplets with a 
gauged shift-symmetry. Other types of models were subsequently discussed in~\cite{Quigley:2011pv,Blaszczyk:2011ib} where logarithmic 
axial couplings of ordinary charged chiral multiplets were used for the same purpose. Latter on it was 
realized~\cite{Blaszczyk:2011ib,Quigley:2011pv,Quigley:2012gq} that these logarithmic terms can arise in certain branches of 
ordinary $(0,2)$ GLSMs, at one loop; thus, it seems that torsional GLSMs with log couplings are ordinary GLSMs in disguise. 
These results motivated a more general study of $(0,2)$ gauge theories initiated in~\cite{Melnikov:2012nm}. 

Mirror symmetry~\cite{Hori:2000kt} can certainly help in understanding the intimate relations between torsional and torsion-free 
compactifications, as it relates generically worldsheet theories whose target spaces have different topologies. Generalization of 
mirror symmetry to $(0,2)$ models was considered first in~\cite{Adams:2003zy} and is not yet well-understood  
(see~\cite{Melnikov:2012hk} for recent developments); including torsional GLSMs has not been done yet.

In this note, as a modest step in this direction, we investigate dualities between the gauged linear 
sigma-models with torsion corresponding to principal two-torus bundles over K3, equipped with some holomorphic gauge bundle 
$V$, and $(0,2)$ GLSMs for $T^2 \times K3$ with an additional line bundle over the K3 surface. Perturbative dualities of this sort have 
been already investigated from a target-space perspective~\cite{Evslin:2008zm,Andriot:2009fp,Martelli:2010jx}, 
using the heterotic generalization of Buscher rules (for a review see~\cite{Giveon:1994fu}). It allowed to relate 
these seemingly distinct class of string backgrounds, however only at lowest order in $\alpha'$; this 
is not really satisfactory in the present context as these solutions involve typically compactification at 
string scale.\footnote{In~\cite{Becker:2007ea,Sethi:2007bw}, 
using an $\mathcal{M}$-theory construction,  it was argued independently that there was a transition between these two types 
of compactifications at orbifold points of the K3 surface.} It was also argued in~\cite{Andriot:2011iw} that dualities 
of this sort allow to embed heterotic solutions in a higher-dimensional theory.

Dualities mapping the torus fiber onto itself are also worthwhile to consider. We provide the mapping of parameters between the 
dual models, and show that quantization constraints on the moduli arise by demanding consistency of the dual theory 
for generic duality transformations. Interestingly, these conditions force the torus moduli to be those of a rational conformal 
field theory.

We will implement these two classes of dualities at the level of the gauged linear sigma-model, showing that 
they are exact symmetries of the heterotic strings; some preliminary steps in this direction can be 
found in~\cite{Adams:2009av}. These dualities are actually symmetries of the GLSM themselves, even though the 
models could, in principle, be distinguished by irrelevant couplings that would not  spoil the dualities between 
their respective infrared fixed points. Technically, we prove these dualities by introducing an extra worldsheet gauge field 
together with an extra chiral superfield, playing the role of a Lagrange multiplier, 
following closely the methods of~\cite{Rocek:1991ps}.\footnote{The same methods can be applied to the solvable 
worldsheet descriptions of non-compact models of this sort, found in~\cite{Carlevaro:2008qf}, 
allowing to prove directly T-duality for superconformal theories.}  For dualities mixing the torus and gauge bundles, 
we point out a subtlety related to the left-moving GSO projection, that would be hard to guess from a supergravity 
perspective. We also discuss the Wilson line moduli and their mixing with torus moduli.

Heterotic compactifications with line bundles attracted lot of attention recently, since they allow for computationally-efficient ways 
of getting standard-model-like spectra, the Hermitian-Yang-Mills equations being simple to solve~\cite{Anderson:2011ns}; they 
also appear naturally in GLSM descriptions of resolved heterotic orbifolds~\cite{Blaszczyk:2011hs}. Hence, being able 
to map them to torsional solutions, consisting in principal torus bundles over a Calabi-Yau base, may shed new light 
on their properties and allow to embed them in a web of dualities involving type IIB or $\mathcal{M}$-theory compactification 
with fluxes. 

This letter is organized as follows. In section~\ref{sec:review} we review the construction of $(0,2)$ GLSMs 
for non-K\"ahler principal bundles over CY bases and make some important remarks about 
moduli quantization. In section~\ref{sec:self} we study self-dualities between models within this category. 
In section~\ref{sec:bund} we obtain more general perturbative dualities, 
mixing the torus bundle with the gauge bundle. Finally in section~\ref{sec:conc} we discuss some implications 
of these results. Useful material is gathered in the appendices.

%%%%%%%%%%%%%%%%%%%%%%%%%%%%%%%%%%%%%%%%%%%%%%%%%%%%%%%%%%%%%%%%%%%%%%%%%%%%%%%%%%%%%%%%%%%%%%%%%%%%%%%%%%%%%%%%%%%%%%%%%%%%%%%%%%%%%%%%%%%%%%%%%%%%%%%%%%%%%%%%%%%

\section{Linear sigma-models for torsional compactifications}
\label{sec:review}
In this section we summarize briefly the construction of gauged linear sigma-models with torsion discovered in~\cite{Adams:2006kb}, 
highlighting the most salient features for our present purposes, and discuss in detail the quantization of 
the torus moduli. We shall first review the geometry of the torsional compactifications that they correspond to.

\subsection{Torsional torus bundles in heterotic supergravity}
Heterotic compactifications to four-dimensions with $\mathcal{N}=2$ space-time supersymmetry are given by solutions of the type 
first obtained by Dasgupta, Rajesh, and Sethi using non-perturbative dualities~\cite{Dasgupta:1999ss} and discussed afterwards 
by many authors. They are given by principal two-torus bundles over a K3 base with Hermitian-Yang-Mills gauge bundles.  

Let us discuss these  $T^2 \hookrightarrow \mathcal{C}_6 \stackrel{\pi}{\to} K3$ solutions explicitly. In our conventions, 
$U$ and $T$ correspond respectively to the generalized K\"ahler structure and to the complex structure, see app.~\ref{apptor}. 
Thereafter we work in $\alpha'=1$ units, $i.e.$ such that the self-dual radius is equal to one. 
One considers a ten-dimensional string-frame metric of the form 
\begin{equation}
\label{metric}
\di s^2 = \eta_{\mu \nu} \di X^\mu \di X^\nu +
\frac{U_2}{T_2} \left|\di x^1 + T \di x^2 +  \pi^\star \, \mathfrak{w}^n\alpha_n \right|^2 
+ e^{2 A(y)} \di s^2 (\mathcal{S})\, ,
\end{equation}
where $\di s^2 (\mathcal{S})$ is a Ricci-flat metric on a K3 surface $\mathcal{S}$ and $e^{2A}$ is 
a warp factor depending on the K3 coordinates only. 

The torus fibration data is given by a set of complex topological charges $\mathfrak{w}^n$ belonging 
to the lattice $\mathbb{Z} + T \, \mathbb{Z}$ and a corresponding basis of anti-self-dual two-forms on K3, that 
we write locally as $\Pi_n =\frac{1}{2\pi}\di \alpha_n$, such that 
\begin{equation}
\Pi_n  \in H^{2}(\mathcal{S},\mathbb{Z}) \cap H^{1,1} (\mathcal{S})  \, .
\label{formcond}
\end{equation}
The two-form $\Pi=\mathfrak{w}^n \Pi_n$ should be primitive w.r.t. the base, namely $J\wedge \Pi=0$, where $J$ is the 
K\"ahler form of K3. The two-forms $\Pi_n$ are, by construction, elements of the Picard group $\text{Pic}(\mathcal{S})$. This 
requires some assumptions on the complex structure of $\mathcal{S}$, since generically the Picard group is empty.\footnote{
$\mathcal{N}=2$ supersymmetry 
allows actually a slightly more general construction. One can take a  
one-form $\theta=\di x^1 + T \di x^2 + \pi^\star \alpha$ where  
$\tfrac{1}{2\pi}\di \alpha =  \Pi_1 + T\, \Pi_2$ is an anti-self-dual $(1,1)$ form 
with $\Pi_{1,2} \in H^2 (\mathcal{S},\mathbb{Z})$.} Supergravity solutions with $\mathcal{N}=1$ supersymmetry in four dimensions can 
be obtained if some $\Pi_n$'s have a $(2,0)$ component (see $e.g.$~\cite{Becker:2009zx}), however there is no known GLSM for such compactifications.

The NS-NS three-form follows then from supersymmetry. Introducing the globally defined one-form 
$\theta = \di x^1 + T \di x^2 +  \pi^\star \, \mathfrak{w}^n\alpha_n$ the torsion is given by
\begin{equation}
H = \star_{\textsc{k3}}\,\di\,  e^{2A} -\frac{U_2}{T_2}\, \text{Re} \, \left( \bar{\theta} \wedge \star_{\textsc{k3}} \di \theta \right)\, .
\end{equation}

A supersymmetric gauge bundle is obtained as the pullback of a holomorphic gauge bundle $V$ on K3 satisfying the zero-slope limit of 
the Hermitian-Yang-Mills equation, namely
\begin{equation}
\label{hym}
F^{2,0} = F^{0,2}= 0 \ , \quad J \lrcorner F =0\, .
\end{equation}
On K3 it implies anti-selfduality, $i.e.$ that the bundle $V$ corresponds to an anti-instanton background.  For line bundles, once imposing 
Dirac quantization for the magnetic flux, these conditions are similar to those satisfied by the two-forms $\Pi_n$ specifying the 
principal torus bundle.

In addition, it is possible to add an Abelian gauge potential over the total space $\mathcal{C}_6$, that would reduce 
to Wilson lines for a product manifold $K3 \times T^2$:
\begin{equation}
\label{connecw}
A =  \mathfrak{T}^I\, \text{Re}\, (\bar{\alpha}^I \theta )\, ,
\end{equation}
parametrized by $\alpha^I = \alpha^{I}_1 + T \alpha^{I}_2$ and 
whose embedding in the commutant of the structure group of $V$ is specified by the commuting matrices $\mathfrak{T}^I$.

As was shown by Fu and Yau~\cite{Fu:2006vj}, taking a solution~(\ref{metric}-\ref{hym}) of 
the $\mathcal{N}=2$ supersymmetry conditions, and provided that the following tadpole condition holds
\begin{equation}
\label{tadpole}
\frac{U_2}{T_2} \int_{\textsc{k3}} \star_\textsc{k3} \, \Pi \wedge \bar{\Pi} - \text{ch}_2 (V) =24\, , 
\end{equation}
there exists a solution of the Bianchi identity for the warp factor $\exp 2A$, which is a function of the base coordinates.\footnote{
Different choices of connection with torsion lead to different differential equations for the 
warp factor; some choices give simpler answers than others~\cite{Becker:2009df}.}

\subsection{(0,2) Gauged Linear Sigma-Models for K3}
We start with an old-fashioned $(0,2)$ gauged linear sigma-model whose infrared fixed point would be 
a heterotic non-linear sigma model with a K3 surface target-space, if an important consistency condition was not blatantly violated. 

A $(0,2)$ GLSM is an Abelian $U(1)^r$ two-dimensional supersymmetric gauge theory coupled to charged chiral and Fermi multiplets. 
For simplicity of presentation we choose a single $U(1)$ gauge group; the generalization to $U(1)^r$ is straightforward and 
will be given in places. We refer to appendix~\ref{app:fields} for superspace conventions and~\ref{app:lag} for the component expansion 
of the Lagrangian.

The GLSM contains a set of chiral superfields $\Phi_I$ of gauge charges $Q_I$, a set of Fermi multiplets $\Gamma_a$ of charges $q_a$ 
and the pair of gauge superfields $(\mathcal{A},\mathcal{V} )$ that is needed to formulate gauge theories in $(0,2)$ superspace. 
The full Lagrangian of the theory is of the generic form (summation over indices $I$ and $a$ is implied):
\begin{multline}
\label{baselac}
\mathcal{L}_{K3} = -\frac{i}{2} \int \di^2 \theta^+ \,  \bar{\Phi}_I \mathcal{D}_- \Phi_I
-\frac12 \int \di^2 \theta^+ \, \bar{\Gamma}_a \Gamma_a 
-\frac{1}{8e^2} \int \di^2 \theta^+ \, \bar{\Upsilon} \Upsilon\\
- \frac{\mu}{2}  \int \di \theta^+ \, 
\Gamma_a J^a (\Phi_I)  + \frac{t}{4}  \int \di \theta^+ \, 
\Upsilon + h.c.
\end{multline}
where the gauge superfield-strength $\Upsilon$, which is a chiral superfield is given in eq.~(\ref{upsidef}). The holomorphic 
functions $J^a (\Phi^I)$ in the chiral superfields play a role analogous to the superpotentials of more 
familiar $(2,2)$ theories. 

A model specified by the Lagrangian~(\ref{baselac}) is gauge-anomalous for a generic choice of charges, since left- and right-handed 
fermions belong to different multiplets. Under a super-gauge-transfor\-mation, 
parametrized by a chiral superfield $\Xi$, the effective Lagrangian is shifted by\footnote{Our convention is $\mathcal{S}_{eff} = 
\frac{1}{2\pi} \int d^2 x\, \mathcal{L}_{eff}$.} 
\begin{equation}
\delta_\Xi \mathcal{L}_{eff} = \frac{\mathfrak{A}}{8}\int \di \theta^+\, \Xi\,  \Upsilon \,  + h.c.\, , \quad \text{with} \qquad 
\mathfrak{A}=Q_i Q^i - q_n q^n \, .
\label{anomalyeqn}
\end{equation}
Well-behaved ordinary GLSMs require the vanishing of this anomaly. In contrast, to formulate torsional GLSMs  we shall be happy with any 
positive value of $\mathfrak{A}$, since~(\ref{anomalyeqn}) will eventually be cancelled by the gauge shift of the torus fibre 
Lagrangian.\footnote{We do also require, in the full theory, the existence of non-anomalous global chiral symmetries, giving  
the right $U(1)$ R-current and the left $U(1)$ flavour-current, that will be respectively part of the infrared superconformal 
algebra and used in the GSO projection, since we want to obtain as an infrared fixed point a 
NSLM for a supersymmetric heterotic string compactification.}

For illustrative purposes, let us consider a realization of  K3  as a quartic in $\mathbb{P}^3$. In order to describe 
the geometry of the compactification one takes four chiral superfields $\Phi_i$ of gauge charge $Q=1$ and a single 
Fermi superfield $\Gamma_0$ of charge $-4$. These charges satisfy $\sum_i Q_i = Q_{\Gamma_0}$, which translates 
geometrically in the CY condition $\text{c}_1 (T)=0$. Associated with these superfields is a superpotential coupling of the form
\begin{equation}
\mathcal{L}_{s} =   \int \di \theta^+ \,  \Gamma_0\, G(\Phi_{i}) + h.c.\, ,
\end{equation}
where $G(\phi_i)=0$ is a quartic that defines the K3 surface carved out of the vacuum manifold $\mathbb{P}^3$ in the 
CY phase $r\gg 1$. The right-handed fermions are, as usual, sections of the tangent bundle. 

Supplementing the K3 surface with a gauge bundle can be done in many ways.  One considers for instance 
an extra set of $s+1$ Fermi multiplets $\Gamma_a$ of charge $q_a=+1$, satisfying the standard chirality constraint 
($i.e.$ $\bar{\mathfrak{D}}_+ \Gamma_a=0$), and a chiral multiplet $P$ of charge $Q_p=-(s+1)$. 
The superpotential couplings associated with these multiplets take the form:
\begin{equation}
\mathcal{L}_w =   \int \di \theta^+ \,  P \, \Gamma_a  J^a (\Phi_{1,\ldots,4}) + h.c.
\end{equation}
where the holomorphic functions $J^a (\phi_i)$ are homogeneous polynomials of degree $s$, as requested in order to ensure 
classical gauge invariance. 

In the CY phase, the bottom components of the Fermi multiplets are sections of (the pullback of) the rank $s$ holomorphic 
vector bundle $\mathcal{V}$ over K3 determined by the short exact sequence
\begin{equation}
0 \longrightarrow \mathcal{V} \longrightarrow \bigoplus_{a=1}^{s+1} \mathcal{O} (1) \stackrel{J^a}{\longrightarrow} 
\mathcal{O} (s+1) \longrightarrow 0\, .
\end{equation}

Since we did not impose the constraint $\mathfrak{A}=0$ for gauge anomaly-cancellation, the model hitherto 
defined does not make sense as a quantum theory.

\subsection{Torus fibration and torsion} 
\label{subsec:fibr}
The main idea of the work~\cite{Adams:2006kb} is to realize the Green-Schwarz mechanism at the gauged linear sigma-model level 
by canceling the quantum gauge non-invariance of the K3 GLSM against the classical gauge variation of another term in the action that 
will eventually describe the fiber of the principal $T^2$ bundle. We consider a generic two-torus whose moduli are 
given by the complex parameters $U$ and $T$, respectively the generalized K\"ahler structure and the complex structure, 
see app.~\ref{apptor} for our conventions. 

For this purpose we add a pair of chiral superfields $\Omega_{1,2}$, that complexify the two-torus coordinates, 
whose imaginary shift-symmetry is gauged. Explicit expressions in components are given in appendix~\ref{app:fields}. 
The scalar fields of these multiplets parametrize a 
$\mathbb{C}^\star\times \mathbb{C}^\star$ that we wish to reorganize as $\mathbb{C}\times T^2$, 
by a change of complex structure in target space, in order to decouple the non-compact directions. 

Generalizing the construction of~\cite{Adams:2006kb} to generic torus moduli, including  
a constant B-field along the fiber ($B=U_1 \di x^1 \wedge \di x^2$), 
the Lagrangian associated with these chiral multiplets reads
\begin{multline}
\label{generigfib}
\allowdisplaybreaks
L_{f}=-\frac{iU_2}{8T_2}  \int \di^2 \theta \left( \Omega_1 + \bar \Omega_1
+ T_1 (\Omega_2 + \bar \Omega_2)+2 (\mathfrak{w}_1 + T_1\mathfrak{w}_2)  \mathcal{A} \right)\ \times \\  \times \ 
\left(\partial_- (\Omega_1 - \bar \Omega_1
+ T_1 (\Omega_2 - \bar \Omega_2))+ 2i(\mathfrak{w}_1+T_1 \mathfrak{w}_2) \mathcal{V}  \right) \\
-\frac{i}{8} U_2 T_2  \int \di^2 \theta \left( \Omega_2 + \bar \Omega_2+2\mathfrak{w}_2  \mathcal{A} \right)
\left(\partial_- (\Omega_2 - \bar \Omega_2)+ 2i\mathfrak{w}_2 \mathcal{V}  \right)\\
+\frac{i}{8}U_1  \int \di^2 \theta \Big[ (\Omega_1 + \bar \Omega_1 +2 \mathfrak{w}_1 \mathcal{A}) 
\left(\partial_- (\Omega_2 - \bar \Omega_2) + 2i\mathfrak{w}_2\mathcal{V}  \right) - \\
\qquad \qquad \qquad \qquad -  (\Omega_2 + \bar \Omega_2 +2 \mathfrak{w}_2 \mathcal{A}) 
\left(\partial_- (\Omega_1 - \bar \Omega_1) + 2i\mathfrak{w}_1\mathcal{V}  \right) 
\Big]\\
-\frac{ih^\ell}{4}  \int \di \theta^+ \, \Upsilon\, \Omega_\ell + h.c. 
\end{multline}
where the shift-symmetry of $\text{Im} (\omega_\ell)$ is gauged, with integer charges $\mathfrak{w}_\ell$. 
At the same time,  the chiral superfields $\Omega_\ell$ are coupled axially to the gauge superfields, through the 
Fayet-Iliopoulos (FI) field-dependent term  that appears the last line of eq.~(\ref{generigfib}). 
The axial couplings $h^\ell$ need to be integers, in order for the action to be invariant under 
$\omega_\ell \sim \omega_\ell + 2 i\pi$ for any value of the two-dimensional instanton number $n=-\frac{1}{2\pi}\int F$. 

The Lagrangian given by eq.~(\ref{generigfib}) is classically non gauge-invariant, because of the axial coupling. Under a 
super-gauge variation, whose parameter is a chiral superfield $\Xi$, the Lagrangian is shifted by
\begin{equation}
\label{gaugevarfib}
\delta_{\Xi} \mathcal{L}_g = \frac{1}{4}   h^\ell \mathfrak{w}_\ell  \int \di \theta^+ \, \Xi\, \Upsilon   + h.c.
\end{equation}
This classical gauge variation can be used to precisely cancel the gauge anomaly~(\ref{anomalyeqn}) of the K3 GLSM, hence implementing 
the Green-Schwartz mechanism on the worldsheet. The values of the couplings $h_\ell$ are then 
determined by quantum gauge-invariance of the whole theory:
\begin{equation}
\label{anomalycond}
2h^\ell \mathfrak{w}_\ell =q_n q^n-Q_i Q^i \, .
\end{equation}
This model can be generalized slightly by allowing Wilson lines; we defer this discussion to section~\ref{sec:bund}.

As it stands, this model describes a non-compact principal bundle over K3. In order to decouple the non-compact part of the 
fiber, one needs to cancel the couplings between the fermion field $\mu_-$ of the 
gauge multiplet and the fermions $\chi_{+, \ell}$ of the shift superfields $\Omega_\ell$. Otherwise, the supersymmetry variation 
of the former $\delta_\epsilon \mu_- = i\epsilon (D+2iF_{01})$ would prevent the decoupling from being compatible with  
worldsheet supersymmetry.

Since these fermionic interactions come from the terms proportional to  $\mathcal{V}   \Omega_\ell $ in superspace,  
one simply needs to choose the parameters of the model in order that these terms vanish altogether. Explicitly it amounts to set
\begin{subequations}
\label{chiralcondit}
\begin{align}
\frac{U_2}{T_2} (\mathfrak{w}_1+ T_1 \mathfrak{w}_2)- U_1  \mathfrak{w}_2 +h^1 &=0\, ,\\
\frac{U_2}{T_2} \left[ (\mathfrak{w}_1+ T_1 \mathfrak{w}_2)T_1+ T_2^{\, 2} \mathfrak{w}_2\right]+ U_1  \mathfrak{w}_1+h^2 &=0\, .
\end{align}
\end{subequations}
Then the anomaly cancellation condition of eq.~(\ref{anomalycond}) can be rewritten as
\begin{equation}
Q_i Q^i = \frac{2 U_2}{T_2} |\mathfrak{w}|^2 +q_n q^n\, ,
\end{equation}
reproducing the integrated Bianchi identity~(\ref{tadpole}). One notices that the contribution from the torus fibration 
is of the same sign as the Fermi multiplets anomaly,  which will fit neatly with the perturbative dualities studied 
in the next sections.  As was explained in~\cite{Evslin:2008zm}, the integrated Bianchi identity, which is a measure of the 
five-brane charge, is indeed a natural T-duality invariant for this class of solutions, as far as dualities along the fivebrane 
worldvolume coordinates, that include the two-torus, are concerned.

Integrating out classically the massive gauge fields, one finds as expected a  
$T^2 \hookrightarrow \mathcal{C}_6 \to K3$ geometry  with non-zero torsion, with $\mathcal{N}=2$ spacetime supersymmetry~\cite{Adams:2006kb}. 
Naturally, quantum corrections need to be taken into account in order to get the superconformal 
non-linear sigma-model at the infrared fixed point.

For future reference let us simplify the Lagrangian   associated with~(\ref{generigfib}), using 
the conditions~(\ref{chiralcondit}) and eq.~(\ref{fullsuper}):
\begin{equation}
\label{fibsimp}
L_{fib} =  L_{free} +  \frac{U_2}{T_2}\int \di^2 \theta^+\, \left[|\mathfrak{m}|^2 \mathcal{A}\mathcal{V}  
- \frac{i}{2}\mathcal{A} \left( \text{Re}(\mathfrak{m})  
\partial_- (\Omega_1 -\bar{\Omega}_1)
+\text{Re}(T^\star\mathfrak{m}) \partial_- (\Omega_2 - \bar \Omega_2)\, 
\right)\right]\, .
\end{equation}
This expression makes clear that the coupling between the gauge superfields and the chiral superfields is only 
left-moving, since only terms in $\partial_- \omega_\ell$ appear in the $\theta^+ \bar{\theta}^+$  component. 

Adding to the base GLSM Lagrangian~(\ref{baselac})  the fiber Lagrangian~(\ref{fibsimp}) defines a consistent quantum theory; it 
is generically possible to arrange the global charges of the fields to get non-anomalous chiral $U(1)_L$ and $U(1)_R$~\cite{Adams:2009zg}, 
which are necessary conditions to get a superconformal fixed point with an appropriate GSO projection.

\subsection{Moduli quantization}
\label{subsec:modul}
The relations~(\ref{chiralcondit}) can be seen as  quantization conditions for the torus moduli $U$ and $T$. Defining the 
complex charge 
\begin{equation}
\mathfrak{w}=\mathfrak{w}_1+ T \mathfrak{w}_2\, ,
\end{equation} 
one finds the constraints:
\begin{subequations}
\label{quantcondi}
\begin{align}
\frac{U_2}{T_2}& \text{Re} (\mathfrak{w}) - U_1 \mathfrak{w}_2 \in \mathbb{Z}\, , \label{quantcondi1} \\
\frac{U_2}{T_2}& \text{Re} (T^\star \mathfrak{w}) + U_1 \mathfrak{w}_1 \in \mathbb{Z}\, .  \label{quantcondi2}
\end{align}
\end{subequations}
Similar conditions arise in supergravity by considering flux quantization, see $e.g.$~\cite{Held:2010az}, and 
were discussed at the GLSM level in~\cite{Adams:2006kb}. However the effect of the constant $B$-field ($i.e.$ of $U_1\neq 0$) was 
not taken into account there. As we shall see below, this $B$-field --~that arises naturally by T-dualizing a 
non-orthogonal torus fibration~-- plays an important role in describing the moduli space of allowed torus fibration over K3.

\subsubsection*{Moduli space of torus fibers}
In torsional GLSMs with a single $U(1)$ gauge group, the quantization conditions~(\ref{quantcondi}) reduce the complex two-dimensional 
moduli space parametrized by $(U,T)$ to a one-dimensional subspace, which is described as follows. Under an arbitrary 
infinitesimal complex structure deformation $\delta T$, the K\"ahler structure has to change as well, 
in order to preserve the conditions~(\ref{quantcondi}). The curve defined by the intersection of the two hypersurfaces 
corresponding to~(\ref{chiralcondit}) is locally given by 
\begin{equation}
\label{utrel}
\delta U = - \frac{U_2}{T_2}\, \frac{\mathfrak{w}}{\bar{\mathfrak{w}}} \times \overline{\delta T}\, .
\end{equation}

More generic models are just as easy to consider. One starts with a K3 GLSM with a $U(1)^r$ gauge symmetry, and a complexified 
two-torus fiber with a set of complex charges $\mathfrak{w}^n$, $n=1,\ldots,r$. For each of these charges one finds quantization 
conditions of the form~(\ref{quantcondi}). Hence one gets as many relations~(\ref{utrel}) between the complex structure and K\"ahler 
structure deformations of the two-torus. If the complex charges $\mathfrak{w}^n$ are not parallel to each other, the moduli are completely 
frozen to discrete values.  As we shall see below, covariance under perturbative dualities brings a complementary viewpoint on this topic.

\subsubsection*{Constraints from dualities}
Let us consider the elliptic curve $E_T=\mathbb{C}/(\mathbb{Z}+ T\mathbb{Z})$ associated with the complex structure of the 
two-torus fiber. Anticipating on the results of section~\ref{sec:self}, by imposing invariance of the theory under T-duality 
we obtain that $\mathfrak{w}\bar{U}$ should belong to the same charge lattice as $\mathfrak{w}$. Assuming 
that this condition should hold  for {\it any} choice of complex topological charge $\mathfrak{w}$, it gives actually a rather 
strong constraint.

Setting apart the trivial (and unphysical) cases with $\bar{U} \in \mathbb{Z}$, this condition means that the elliptic curve $E_T$ should 
admit a non-trivial endomorphism:
\begin{equation}
\phi \ : \quad 
\begin{array}{lcl} E_T &\to& E_T\\
z & \mapsto & \bar{U} z 
\end{array}\, .
\end{equation}
This property, known as {\it complex multiplication}, it only shared by elliptic curves whose complex structure $T$
is valued in an {\it imaginary quadratic number field} $\mathbb{Q}(\sqrt{D})$, namely
\begin{equation}
\label{imquadef}
T \in \mathbb{Q} +\sqrt{D}\, \mathbb{Q}\quad \text{with} 
\qquad D=b^2-4ac<0 \quad , \qquad a,b,c \in \mathbb{Z} \, .
\end{equation}
For a nice presentation of these topics from a string theory perspective, see~\cite{Moore:2004fg}.

To be more explicit, the condition for $\phi$ to be an automorphism of $E_T$ can be solved with the sufficient conditions
\begin{equation}
\label{imquad}
\left\{ 
\begin{array}{ll}
\exists\, (m_1,n_1) \in \mathbb{Z}^2 \, , \ \bar U = m_1 + T n_1\\[4pt]
\exists\, (m_2,n_2) \in \mathbb{Z}^2 \, , \  \bar U T = m_2 + T n_2
\end{array}\right.
 \, .
\end{equation}
These relations lead to a second-order equation for the complex structure $T$ with integer coefficients, hence 
$T$ belongs to some $\mathbb{Q} (\sqrt{D})$. Furthermore, eq.~(\ref{imquad}) implies that both $U$ and $T$ are valued 
in the same imaginary quadratic field.  Whenever this is the case, the quantization conditions~(\ref{quantcondi}) are easily solved 
since they now involve only rational numbers.\footnote{Likewise, under T-duality along $x^1$, which exchanges 
the complex and K\"ahler structures, one has the map $\mathfrak{w}\, \mapsto\, \frac{U_2}{T_2} \mathfrak{w}$.
Therefore, as the dual charge has to belong to the lattice $\mathbb{Z}+U\mathbb{Z}$, one has the constraint 
\begin{equation*}
\exists\, (r_1,r_2)\, \in \, \mathbb{Z}^2 \ , \quad \frac{U_2}{T_2} (\mathfrak{w}_1 + T \mathfrak{w}_2 ) = r_1 + U r_2 \, ,
\end{equation*}
which is automatically satisfied using the quantization condition~(\ref{quantcondi1}).}

Interestingly, a two-dimensional conformal field theory with a two-torus target space whose complex structure $T$ and K\"ahler 
structure $U$ are valued in the same imaginary quadratic number field $\mathbb{Q} (\sqrt{D})$ is a rational conformal field 
theory~\cite{Gukov:2002nw}, $i.e.$ with an extended chiral algebra with respect to which the number of primary fields is of 
finite number.\footnote{When the base is an Eguchi-Hanson space, in which case a worldsheet CFT is known, this 
requirement is indeed absolutely necessary and quite easy to understand~\cite{Carlevaro:2011mn}. 
The existence of chiral generators of an extended algebra is 
needed in order to define the super-Liouville potential that appears in these constructions.} 
This peculiar quantization of the moduli fits neatly with known facts about string duals of the heterotic solutions under 
consideration, as we shall see below.

Let us consider a type IIB orientifold $K3 \times 
T^2/\mathbb{Z}_2$ with flux~\cite{Dasgupta:1999ss}, where the orientifold action is defined as 
$\mathbb{Z}_2 = \Omega(-)^{F_L} \mathcal{I}$ with $\mathcal{I}$ the 
inversion along the torus; we take a square torus for simpli\-city. 
The type IIB string-frame metric for the compact space is of the form 
(we use thereafter tilded variables to denote type IIB quantities)
\begin{equation}
\label{iibmetric}
\di \tilde{s}^2 =  \Delta  \left[ \tilde{R}_1^{\,2 }(\di \tilde{x}^1)^{\,2} + \tilde{R}_2^{\, 2}
(\di \tilde{x}^2)^{\, 2}   + \di s^2 (K3) \right]\, ,
\end{equation}
where the overall warp factor $\Delta$ is a function of the K3 coordinates. 
As explained in~\cite{Tripathy:2002qw,Moore:2004fg}, $\mathcal{N}=2$-preserving fluxes fix  neither the complex 
structure of K3 nor the K\"ahler structure of the $T^2$ completely, but constrains the complex structure of the torus 
$\tilde{T}$ and the axio-dilaton $\tilde{\Phi} = a + i/\tilde{g}_s$, which is constant in this type IIB background, 
to be valued in the same imaginary quadratic number field. 

Mapping this type IIB orientifold compactification first to  type I (by using the $U\to -1/U$ and $T\to -1/T$ elements of 
the toroidal T-duality group along the two-torus), then to a $Spin(32)/\mathbb{Z}_2$ heterotic solution by using S-duality, 
leads to the string-frame metric
\begin{equation}
\di s^2 = 
\frac{1}{\tilde{g}_s }\frac{\tilde{R}_2}{\tilde{R}_1}
\left[(\di x^1+\pi^\star \mathfrak{w}_{1,\ell} \alpha^{\ell})^2 + 
\left(\frac{\tilde{R}_1}{\tilde{R}_2}\right)^2 
(\di x^2+\pi^\star \mathfrak{w}_{2,\ell} \alpha^{\ell})^2 \right]
+  \frac{\tilde{R}_1 \tilde{R}_2}{\tilde{g}_s}\Delta^2 \, \di s^2 (K3)\, . 
\end{equation}
The torus is now fibered over K3, as follows from T-dualizing the NSNS two-form of the type IIB solution. 

This heterotic background is of the same type as those given by eq.~(\ref{metric}), 
upon identifying the warp factor as $e^{2A} = \frac{\tilde{R}_1 \tilde{R}_2}{\tilde{g}_s}\Delta^2$ and the torus moduli as 
$T_2 = \tilde{R}_1/\tilde{R}_2 = 1/\tilde{T}_2$ and $U_2= 1/\tilde{g}_2$.  More generally, the chain of dualities relates  
the type IIB and heterotic moduli as follows (up to a $PSL(2,\mathbb{Z})_T$ transformation):
\begin{equation}
T_\textsc{het} = T_\textsc{iib} \quad , \qquad U_\textsc{het} = \Phi_\textsc{iib}\, ,
\end{equation}
hence mapping the quantization conditions for the axio-dilaton and torus complex structure found in type IIB 
to the quantization conditions for both torus moduli in heterotic, that we have found above.

To take another point of view, heterotic strings on $K3\times T^2$, with the two-torus at a rational conformal field 
theory point ($i.e.$ with $T,U\in \mathbb{Q}(\sqrt{D})$ as we have seen) is dual to $\mathcal{F}$-theory on 
$K_3\times K3'$, where the  $K3'$ surface 
is of a special type, called {\it attractive} by Greg Moore~\cite{Moore:2004fg}, having its 
Picard group of maximal dimension. We refer to this reference for more details about their properties. 
Adding flux to these $\mathcal{F}$-theory compactifications allows one to get the torsional 
solutions of interest in heterotic. Conditions imposed by G-flux have been analysed in the $\mathcal{M}$-theory 
dual~\cite{Dasgupta:1999ss}, and precisely single out attractive K3 surfaces.

%%%%%%%%%%%%%%%%%%%%%%%%%%%%%%%%%%%%%%%%%%%%%%%%%%%%%%%%%5

\section{Torus self-duality}
\label{sec:self}
In this section we study the action of perturbative dualities along the two-torus fiber.
In order to derive these dualities {\it \`a la} Buscher from the worldsheet theory, we add an extra $U(1)$ gauge 
symmetry (denoted with a hat thereafter)  to the torsional GLSM, corresponding to the pair of 
superfields $(\hat{\mathcal{A}},\hat{\mathcal{V}})$, and gauge the imaginary shift-symmetry of the superfields $\Omega_\ell$. 

\subsection{Warm-up}
For starters, let us consider a free chiral superfield  $\Phi$ whose bottom component parametrizes a cylinder 
$\mathbb{C}^\star \simeq \mathbb{R}\times S^1$, 
with $\phi \sim \phi + 2i\pi$. Gauging the (compact) shift symmetry of the imaginary part of $\phi$, 
we consider the Lagrangian superspace density:
\begin{equation}
\label{freed}
L = -\frac{i R^2}{8} \left( \Phi + \bar \Phi +2 \hat{\mathcal{A}} \right)
\left(\partial_- (\Phi-\bar\Phi)+ 2i\hat{\mathcal{V}} \right) + 
\frac{1}{4}\left(\hat{\mathcal{V}} (\Delta+\bar\Delta) + i \hat{\mathcal{A}} \partial_- (\Delta-\bar \Delta)
\right)\, .
\end{equation}
Integrating out $\Delta$ first yields the equation
\begin{equation}
0=\bar{D}_+ (\partial_- \hat{\mathcal{A}} +i\hat{\mathcal{V}} )=  \hat{\Upsilon}\, , 
\end{equation}
hence giving back the original action.

In order to obtain the dual description, one integrates out first the gauge fields instead. Fixing the   
the gauge with $\Phi=0$, one gets the equations of motion
\begin{equation}
\label{gaugeeom}
R^2 \hat{\mathcal{V}} = - \frac{i}{2} 
\partial_- (\Delta-\bar \Delta) \quad, \qquad R^2 \hat{\mathcal{A}} = - \frac{1}{2}  (\Delta+\bar \Delta)\, .
\end{equation}
Therefore the dual Lagrangian density reads
\begin{equation}
\label{dualfree}
\tilde{L}= - \frac{i}{8 R^2}(\Delta+\bar \Delta)\partial_- (\Delta-\bar \Delta)\, ,
\end{equation}
which is the expected $R\leftrightarrow \tfrac{1}{R}$ toroidal T-duality as far as $\text{Im}\, \phi$ is concerned. 

\subsection*{Minimally vs axially coupled chirals}
It is also possible to dualize a chiral superfield $\Phi$ whose shift-symmetry was already gauged, associated 
with a pair of gauge superfields $(\mathcal{A},\mathcal{V} )$, and with shift-charge $q$. 
One considers the following Lagrangian density in superspace (omitting the kinetic term of $\Upsilon$ that plays no role):
\begin{multline}
L = -\frac{i R^2}{8} \left( \Phi + \bar \Phi +2q \mathcal{A}+2 \hat{\mathcal{A}} \right)
\left(\partial_- (\Phi-\bar\Phi)+2iq\mathcal{V}  + 2i\hat{\mathcal{V}} \right) \\+ 
\frac{1}{4}\left(\hat{\mathcal{V}} (\Delta+\bar\Delta) + i \hat{\mathcal{A}} \partial_- (\Delta-\bar \Delta)
\right)\, .
\end{multline}
Integrating out the gauge superfields $\hat{\mathcal{A}}$ and $\hat{\mathcal{V}}$ in the gauge $\Phi=0$ 
gives the dual model, corresponding to a neutral chiral superfield axially coupled to the gauge 
superfields $(\mathcal{A},\mathcal{V})$ through a FI-like term:
\begin{equation}
\tilde{L} = - \frac{i}{8 R^2}(\Delta+\bar \Delta)\partial_- (\Delta-\bar \Delta) 
- \frac{q}{4} \left(\mathcal{V}  (\Delta+\bar\Delta) + i \mathcal{A} \partial_- (\Delta-\bar \Delta) \right)\, .
\end{equation}

In the opposite situation, one starts with an axially coupled neutral chiral superfield with axial coupling $h\in \mathbb{Z}$:
\begin{equation}
\label{Axialcoupl}
\frac{h}{4}\left(\mathcal{V}  (\Phi+\bar\Phi) + i \mathcal{A} \partial_- (\Phi-\bar \Phi) \right)\, .
\end{equation}
One minimally couples this superfield to the hatted gauge superfields as in eq.~(\ref{freed}). However, the 
axial coupling~(\ref{Axialcoupl}) is not invariant under the newly introduced hatted gauge symmetry. Hence 
the latter should be corrected to 
\begin{equation}
\label{AxialcouplNew}
L_A = \frac{h}{4}\left[\mathcal{V}  (\Phi+\bar\Phi+2\hat{\mathcal{A}}) + 
i \mathcal{A} \left(\partial_- (\Phi-\bar \Phi)  +2i \hat{\mathcal{V}} \right) \right]
\end{equation}
Notice that, under an unhatted gauge transformation, this axial coupling shifts by the term  
\begin{equation}
\label{nongaugeinv}
\delta_{\Xi} L_A = \frac{ih}{4} \Big( \Xi (\hat{\mathcal{V}} -i\partial_- \hat{\mathcal{A}}) +
\bar \Xi (\hat{\mathcal{V}} +i\partial_- \hat{\mathcal{A}} ) \Big)\, ,
\end{equation}
which looks like a gauge anomaly for the hatted $U(1)$ supergauge symmetry but for a gauge transformation of 
the original $U(1)$. In order to compensate for this classical gauge non-invariance, we need to assign a shift charge $-h$ w.r.t. 
the unhatted gauge symmetry to the chiral superfield $\Delta$. Then the 
gauge variation of the coupling
$$
L_\Delta= \frac{1}{4}\left(\hat{\mathcal{V}} (\Delta+\bar\Delta) + i \hat{\mathcal{A}} \partial_- (\Delta-\bar \Delta) \right)
$$
precisely cancels~(\ref{nongaugeinv}). Integrating out the hatted gauge fields in the gauge $\Phi=0$ gives the dual Lagrangian density
\begin{equation}
\tilde{L} = -\frac{i}{8 R^2} \left(\Delta + \bar{\Delta} 
- 2h\mathcal{A}\right)\left(\partial_- (\Delta - \bar{\Delta}) - 2ih \mathcal{V} \right)
\end{equation}
which is indeed a minimally coupled chiral field of shift charge $\tilde{q}=-h$.

\subsection{Dualizing the fibered torus}
We consider now the model of interest, namely a torsional GLSM for a $T^2 \hookrightarrow \mathcal{C}_6 \to K3$ compactification. 
Our aim is to show that this set of compactifications is closed under $O(2,2,\mathbb{Z})$ perturbative dualities along the torus 
fiber. We will derive the transformation rules for the charge doublet $(\mathfrak{w}_1,\mathfrak{w}_2)$ under 
such transformations. Dualities belonging to the larger $O(2,2+N,\mathbb{Z})$ heterotic duality 
group will be considered in the next section.

The K3 GLSM plays no role in the duality, hence we write below only the terms in the Lagrangian density 
associated with the torus fiber for generic moduli $U$ and $T$, see eq.~(\ref{generigfib}). 
In contrast with the previous examples, the chiral superfields are coupled axially {\it and} 
minimally at the same time. Our aim is to find a set of duality transformations generating the full 
$PSL(2,\mathbb{Z}))_T \times PSL(2,\mathbb{Z})_U \times \mathbb{Z}_2 \times \mathbb{Z}_2$ perturbative duality group.

Let us dualize the chiral superfield $\Omega_1$ first, coupling also $\Omega_2$ axially to the hatted gauge 
superfields, with a parameter $s\in \mathbb{Z}$. One starts then with the following superspace Lagrangian density: 
\begin{multline}
\label{dualstart}
%\allowdisplaybreaks
L_s =-\frac{iU_2}{8T_2}  \left( \Omega_1 + \bar \Omega_1
+ T_1 (\Omega_2 + \bar \Omega_2)+2 (\mathfrak{w}_1 + T_1\mathfrak{w}_2)  \mathcal{A}+2\hat{\mathcal{A}} \right)\ \times \\  \times \ 
\left(\partial_- (\Omega_1 - \bar \Omega_1
+ T_1 (\Omega_2 - \bar \Omega_2))+ 2i(\mathfrak{w}_1+T_1 \mathfrak{w}_2) \mathcal{V}  +2i\hat{\mathcal{V}} \right) \\
-\frac{i}{8} U_2 T_2 \left( \Omega_2 + \bar \Omega_2+2\mathfrak{w}_2  \mathcal{A} \right)
\left(\partial_- (\Omega_2 - \bar \Omega_2)+ 2i\mathfrak{w}_2 \mathcal{V}  \right)\\
+\frac{i}{8}U_1  \Big[ (\Omega_1 + \bar \Omega_1 +2 \mathfrak{w}_1 \mathcal{A}+2\hat{\mathcal{A}}) 
\left(\partial_- (\Omega_2 - \bar \Omega_2) + 2i\mathfrak{w}_2\mathcal{V}  \right) -\quad\\ 
\qquad \qquad -  (\Omega_2 + \bar \Omega_2 +2 \mathfrak{w}_2 \mathcal{A}) 
\left(\partial_- (\Omega_1 - \bar \Omega_1) + 2i\mathfrak{w}_1\mathcal{V} +2i\hat{\mathcal{V}}\right) 
\Big]\\
+\frac{h_1}{4} \left[\mathcal{V}  (\Omega_1+\bar\Omega_1+2\hat{\mathcal{A}}) + i \mathcal{A} \left(\partial_- (\Omega_1-\bar\Omega_1)+2i\hat{\mathcal{V}} \right)\right]
+\frac{h_2}{4} \left[\mathcal{V}  (\Omega_2+\bar\Omega_2) + i \mathcal{A} \partial_- (\Omega_2-\bar\Omega_2)\right]\\
+\frac{1}{4} \left[\hat{\mathcal{V}} (\Theta+\bar\Theta) + i \hat{\mathcal{A}}\partial_- (\Theta-\bar\Theta)
\right]
+\frac{s}{4} \left[\hat{\mathcal{V}} (\Omega_2+\bar{\Omega}_2) 
+ i \hat{\mathcal{A}} \partial_- (\Omega_2-\bar{\Omega}_2) 
\right]\, .
\end{multline}
Neither $\Theta$ nor $\Omega_2$ is charged under the hatted gauge symmetry hence gauge invariance w.r.t. the latter is automatic. 

Gauge-invariance w.r.t. the unhatted gauge symmetry, corresponding to the superfields $\mathcal{A}$ and $\mathcal{V} $, 
should be examined more carefully. Assigning a shift-charge $\mathfrak{w}_\theta$ to the chiral superfield $\Theta$, under 
a super-gauge variation, the Lagrangian density~(\ref{dualstart}) is classically shifted by the term (after integration by parts)
\begin{equation}
\delta_{\Xi} L_s = \frac{h^\ell \mathfrak{w}_\ell}{4} \int \di \theta^+\,  \Xi \Upsilon + \frac{1}{4} \left( h_1 + \mathfrak{w}_\theta 
+ s \mathfrak{w}_2 \right) \int \di \theta^+\, \Xi \hat \Upsilon + h.c.
\end{equation}
The first term is, as above, responsible for canceling the gauge anomaly coming from the base GLSM, see eq.~(\ref{anomalycond}). The 
second term should vanish, thereby fixing the value of the $\Theta$ shift-charge to (see the discussion below~(\ref{nongaugeinv})):
\begin{equation}
\mathfrak{w}_\theta = - h_1 - s \mathfrak{w}_2  = \frac{U_2}{T_2} (\mathfrak{w}_1+ T_1 \mathfrak{w}_2)- (U_1+s)  \mathfrak{w}_2 \, ,
\end{equation}
where we used the relation~(\ref{chiralcondit}).

After being reassured that gauge-invariance is satisfied, one can simplify the bulky expression~(\ref{dualstart}) by using 
one more time eq.~(\ref{chiralcondit}) and by fixing the gauge $\Omega_1=0$:
\begin{multline}
L_s =-\frac{i}{8}\frac{U_2}{T_2} |T|^2 (\Omega_2+\bar \Omega_2) \partial_- (\Omega_2 - \bar \Omega_2) 
+ \frac{U_2}{2T_2}|\mathfrak{m}|^2 \mathcal{A}\mathcal{V}  
+ \frac{i}{2}\mathcal{A} \left(h_2+U_1 \mathfrak{w}_1\right) \partial_- (\Omega_2 - \bar \Omega_2)\\
+\frac{U_2}{2T_2}\hat{\mathcal{A}}\hat{\mathcal{V}} 
-(-U_1 \mathfrak{w}_2+h_1) \mathcal{A}\hat{\mathcal{V}} 
 +\frac{1}{4} \left[\hat{\mathcal{V}} (\Theta+\bar\Theta+U_1 (\Omega_2 + \bar \Omega_2)) + i \hat{\mathcal{A}}
\partial_- (\Theta-\bar\Theta+U_1 (\Omega_2 - \bar \Omega_2))\right]\\
+ \frac{1}{4}\frac{U_2 T_1}{T_2}
\left[ \hat{\mathcal{V}} (\Omega_2+\bar{\Omega_2}) - i \hat{\mathcal{A}}\partial_- (\Omega_2-\bar{\Omega}_2)\right]
+\frac{s}{4} \left[\hat{\mathcal{V}} (\Omega_2+\bar{\Omega}_2) 
+ i \hat{\mathcal{A}} \partial_- (\Omega_2-\bar{\Omega}_2) 
\right]\,.
\end{multline}
Now integrating out the hatted gauge superfields gives
\begin{subequations}
\begin{align}
\hat{\mathcal{V}} &=  -\frac{i}{2} \frac{T_2}{U_2} 
\partial_- \Big( \Theta - \bar \Theta + (U_1+s) (\Omega_2 - \bar \Omega_2 ) \Big) 
+ \frac{i}{2} T_1 \partial_- (\Omega_2 - \bar \Omega_2)\, ,\\
\hat{\mathcal{A}} &= - \frac{1}{2} \frac{T_2}{U_2} \Big(\Theta + \bar \Theta 
+ (U_1+s) (\Omega_2 + \bar \Omega_2 )\Big) + \frac{T_2}{U_2} (-2U_1 \mathfrak{w}_2+2h_1)  
\mathcal{A}- \frac{1}{2}T_1(\Omega_2+\bar{\Omega_2})\, ,
\end{align}
\end{subequations}
leading to the dual Lagrangian density:
\begin{multline}
\label{duallagg}
\tilde{L} = -\frac{i}{8} \frac{T_2}{U_2}  \Big(\Theta + \bar \Theta + (U_1+s) (\Omega_2 + \bar \Omega_2 )\Big)  
\partial_-  \Big(\Theta - \bar \Theta + (U_1+s) (\Omega_2 - \bar \Omega_2 )\Big)\\
-\frac{i}{8} U_2 T_2 (\Omega_2+\bar \Omega_2) \partial_- (\Omega_2 - \bar \Omega_2)
-\frac{i}{8} T_1 \left( (\Omega_2+\bar{\Omega}_2) \partial_- (\Theta-\bar \Theta) - (\Theta+\bar{\Theta}) 
\partial_- (\Omega_2-\bar{\Omega}_2) \right)  \\
+ \frac{U_2}{2T_2}\left|\mathfrak{w}\right|^2 \mathcal{A}\mathcal{V}  
-\frac{i}{2}\left[U_2 T_2 \mathfrak{w}_2 + (\mathfrak{w}_1+T_1 \mathfrak{w}_2)(U_1+s) \right]
\mathcal{A}\partial_- (\Omega_2 - \bar \Omega_2)
- \frac{i}{2} (\mathfrak{w}_1+T_1 \mathfrak{w}_2)
\mathcal{A}\partial_- \Big( \Theta - \bar \Theta \Big)\, .\\
\end{multline} 
We read then the duality transformations of the torus moduli from the first two lines of eq.~(\ref{duallagg}):
\begin{equation} 
\mathcal{G}_s \ : \quad \left\{ \begin{array}{ll} U & \mapsto \tilde{U}=T \\
T & \mapsto \tilde{T}=U+s
\end{array} \right. \, .
\end{equation}
Given that the dual Lagrangian is of the same form as the original one, we define 
the dual complex topological charge with the help of the dual complex structure:
\begin{equation}
\tilde{\mathfrak{w}}= \tilde{\mathfrak{w}}_1 + \tilde{T} \tilde{\mathfrak{w}}_2 = 
\tilde{\mathfrak{w}}_1 + (U+s) \tilde{\mathfrak{w}}_2
\end{equation}
Comparing with the generic model, given by eq.~(\ref{generigfib}), we find the following duality map between the charges 
specifying the principal bundle:
\begin{subequations}
\begin{align}
\text{Re} (\tilde{\mathfrak{w}}) &= \frac{U_2}{T_2}\text{Re} (\mathfrak{w}) \, ,\notag \\
\text{Re} \left( (\overline{U}+s) \tilde{\mathfrak{w}}\right) &= \frac{U_2}{T_2} \text{Re} \left((\overline{U}+s)\mathfrak{w}\right)\, .
\end{align}
\end{subequations}

We consider now dualizing the chiral superfield $\Omega_2$ instead. One starts  with the superspace Lagrangian density
\begin{multline}
\label{dualstart2}
%\allowdisplaybreaks
L =-\frac{iU_2}{8T_2}  \left( \Omega_1 + \bar \Omega_1
+ T_1 (\Omega_2 + \bar \Omega_2)+2 (\mathfrak{w}_1 + T_1\mathfrak{w}_2)  \mathcal{A}+2T_1 \hat{\mathcal{A}} \right)\ \times \\  \times \ 
\left(\partial_- (\Omega_1 - \bar \Omega_1
+ T_1 (\Omega_2 - \bar \Omega_2))+ 2i(\mathfrak{w}_1+T_1 \mathfrak{w}_2) \mathcal{V}  +2iT_1\hat{\mathcal{V}} \right) \\
-\frac{i}{8} U_2 T_2 \left( \Omega_2 + \bar \Omega_2+2\mathfrak{w}_2  \mathcal{A} + 2\hat{\mathcal{A}}\right)
\left(\partial_- (\Omega_2 - \bar \Omega_2)+ 2i\mathfrak{w}_2 \mathcal{V}  +2i\hat{\mathcal{V}}\right)\\
+\frac{i}{8}U_1  \Big[ (\Omega_1 + \bar \Omega_1 +2 \mathfrak{w}_1 \mathcal{A}) 
\left(\partial_- (\Omega_2 - \bar \Omega_2) + 2i\mathfrak{w}_2\mathcal{V}  +2i\hat{\mathcal{V}} \right) -\qquad \\
\qquad \qquad -  (\Omega_2 + \bar \Omega_2 +2 \mathfrak{w}_2 \mathcal{A} +2\hat{\mathcal{A}}) 
\left(\partial_- (\Omega_1 - \bar \Omega_1) + 2i\mathfrak{w}_1\mathcal{V} \right) 
\Big]\\
+\frac{h_1}{4} \left[\mathcal{V}  (\Omega_1+\bar\Omega_1) + i \mathcal{A}\partial_- (\Omega_1-\bar\Omega_1)\right]
\\+\frac{h_2}{4} \left[\mathcal{V}  (\Omega_2+\bar\Omega_2+2\hat{\mathcal{A}}) 
+ i \mathcal{A}  \left(\partial_- (\Omega_2-\bar\Omega_2)+2i\hat{\mathcal{V}} \right) \right]
+\frac{1}{4} \left[\hat{\mathcal{V}} (\Theta+\bar\Theta) + i \hat{\mathcal{A}}\partial_- (\Theta-\bar\Theta)
\right]\, .
\end{multline}
Integrating out the hatted gauge fields and taking the gauge $\Omega_2=0$, one is left to the dual Lagrangian 
\begin{multline}
\tilde{L} = - \frac{i}{8} \frac{U_2 T_2}{|T|^2} ( \Omega_1 + \bar \Omega_1) \partial_- (\Omega_1-\bar{\Omega}_1) \\
-\frac{i}{8} \frac{T_2}{U_2 |T|^2} \left(\Theta+\bar \Theta-U_1(\Omega_1+\bar{\Omega}_1)\right)\partial_- \Big(\Theta-\bar \Theta 
- U_1 (\Omega_1-\bar \Omega_1 )\Big)\\
+\frac{i}{8} \frac{T_1}{|T|^2} \left[ (\Theta+\bar \Theta )\partial_- (\Omega_1-\bar{\Omega}_1) -
(\Omega_1+\bar{\Omega}_1) \partial_- (\Theta-\bar \Theta )\right]\\+ \frac{U_2}{2T_2} |\mathfrak{w}|^2 \mathcal{A} \mathcal{V} 
-\frac{i}{2} \frac{\text{Re}(\bar T \mathfrak{w})}{|T|^2} \mathcal{A} \partial_- \Big(\Theta-\bar \Theta \Big)\
+\frac{i}{2} \frac{\text{Re}(\overline{TU} \mathfrak{w})}{|T|^2} \mathcal{A} \partial_- (\Omega_1-\bar{\Omega}_1)\, .
\end{multline}
Hence the torus moduli transform as follows:
\begin{equation} 
\mathcal{H} \ : \quad \left\{ \begin{array}{ll} U & \mapsto \tilde{U}=\frac{1}{\bar T} \\
T & \mapsto \tilde{T}=-\bar U
\end{array} \right. \, .
\end{equation}
Accordingly, defining now $\tilde{\mathfrak{w}}= \tilde{\mathfrak{w}}_1 -\bar U \tilde{\mathfrak{w}}_2$, one has
\begin{subequations}
\begin{align}
\text{Re} (\tilde{\mathfrak{w}}) &= \frac{U_2}{T_2}\text{Re} (\bar T  \mathfrak{w}) \, ,\notag \\
\text{Re} ( U \tilde{\mathfrak{w}}) &= \frac{U_2}{T_2} \text{Re} \left( \overline{TU} \mathfrak{w}\right)\, .
\end{align}
\end{subequations}
Had we had chosen instead the axial coupling of $\Theta$ with an opposite sign, one would end up with 
another dual Lagrangian density
\begin{multline}
\label{Idual}
\tilde{L} = - \frac{i}{8} \frac{U_2 T_2}{|T|^2} ( \Omega_1 + \bar \Omega_1) \partial_- (\Omega_1-\bar{\Omega}_1) \\
-\frac{i}{8} \frac{T_2}{U_2 |T|^2} \left(\Theta+\bar \Theta+U_1(\Omega_1+\bar{\Omega}_1)\right)\partial_- \Big(\Theta-\bar \Theta 
+ U_1 (\Omega_1-\bar \Omega_1 )\Big)\\
-\frac{i}{8} \frac{T_1}{|T|^2} \left[ (\Theta+\bar \Theta )\partial_- (\Omega_1-\bar{\Omega}_1) -
(\Omega_1+\bar{\Omega}_1) \partial_- (\Theta-\bar \Theta )\right]\\+ \frac{U_2}{2T_2} |\mathfrak{w}|^2 \mathcal{A} \mathcal{V} 
+\frac{i}{2} \frac{\text{Re}(\bar T \mathfrak{w})}{|T|^2} \mathcal{A} \partial_- \Big(\Theta-\bar \Theta)\Big)\
+\frac{i}{2} \frac{\text{Re}\left(\overline{TU} \mathfrak{w}\right)}{|T|^2} \mathcal{A} \partial_- (\Omega_1-\bar{\Omega}_1)\, ,
\end{multline}
corresponding to the transformations
\begin{equation} 
\mathcal{I} \ : \quad \left\{ \begin{array}{ll} U & \mapsto \tilde{U}=-\frac{1}{T} \\
T & \mapsto \tilde{T}=U
\end{array} \right. \,.
\end{equation}
As before, defining $\tilde{\mathfrak{w}}= \tilde{\mathfrak{w}}_1 +U \tilde{\mathfrak{w}}_2$ leads to the map
\begin{subequations}
\begin{align}
\text{Re} (\tilde{\mathfrak{w}}) &= -\frac{U_2}{T_2}\text{Re} (\bar T \mathfrak{w})\, , \notag \\
\text{Re} ( \bar U \tilde{\mathfrak{w}}) &= -\frac{U_2}{T_2} \text{Re} \left( \overline{TU} \mathfrak{w}\right)\, .
\end{align}
\end{subequations}

\subsection{Perturbative duality group}
The set of duality transformations that were derived above by expressing them as quotients allows one to generate the full 
$O(2,2,\mathbb{Z})$ perturbative duality group along the two-torus fiber.

First, let us discuss the $PSL(2,\mathbb{Z})_T$ factor, acting on the complex structure. 
The $\mathcal{T}_T$ and $\mathcal{S}_T$ generators are respectively given in terms of the previously 
defined dualities as $\mathcal{G}_1 \mathcal{G}_0$ and $\mathcal{G}_0\mathcal{I}$. 
The corresponding duality transformations for the complex charge $\mathfrak{w}$ are then given by
\begin{equation}
\mathcal{T}_T \, : \mathfrak{w}\mapsto \mathfrak{w} \quad, \qquad \mathcal{S}_T \, : \mathfrak{w}\mapsto -\frac{1}{T}\mathfrak{w}\, .
\end{equation} 
In other words, $(\mathfrak{w}_1,\mathfrak{w}_2)$ transforms as a doublet of $PSL(2,\mathbb{Z})_T$. Under 
$T\to \frac{aT+b}{cT+d}$ one has the map
\begin{equation}
\left( \begin{array}{c} \mathfrak{w}_1 \\ \mathfrak{w}_2 \end{array} \right) \mapsto 
\left( \begin{array}{ccc} a && b  \\ c && d \end{array} \right)
\left( \begin{array}{c} \mathfrak{w}_1 \\ \mathfrak{w}_2 \end{array} \right)\, ,
\end{equation}
which is not a surprising result. 

The $PSL(2,\mathbb{Z})_U$ factor, acting on the K\"ahler structure, is generated by the elements
$\mathcal{T}_U= \mathcal{G}_0 \mathcal{G}_1$ and $\mathcal{S}_U=\mathcal{I}\mathcal{G}_0$. Using the previous results 
one finds that the complex topological charge transforms under these dualities as: 
\begin{equation}
\label{pslu}
\mathcal{T}_U \, : \mathfrak{w}\mapsto \mathfrak{w} \quad, \qquad \mathcal{S}_U \, : \mathfrak{w}\mapsto \, 
-\bar U \mathfrak{w} \, .
\end{equation}
The behavior of $\mathfrak{w}$ under $\mathcal{S}_U$ makes only sense if 
the transformed charge $-\bar U \mathfrak{w}$ belongs to the same charge lattice as $\mathfrak{w}$, namely 
$\mathbb{Z} + T\mathbb{Z}$. Implications of this important result for moduli quantization were discussed 
in subsection~\ref{subsec:modul}.

Finally, a first $\mathbb{Z}_2$ factor, T-duality along $x_1$, is directly given by $\mathcal{G}_0$ with the 
complex charge transforming as $\mathfrak{w} \to \frac{U_2}{T_2} \mathfrak{w}$. The second $\mathbb{Z}_2$, 
namely the parity transformation, corresponds to $ \mathcal{I}\mathcal{G}_0\mathcal{H}$ and 
maps $\mathfrak{w} \to-\bar{\mathfrak{w}}$. This completes the perturbative duality group of the two-torus. 

Hence we have proven that the whole $O(2,2,\mathbb{Z})$ perturbative duality group is an exact symmetry of the class of torsional 
gauged linear sigma-models under consideration, and not only a symmetry among their infrared fixed points.

One can wonder whether non-perturbative effects (a.k.a. worldsheet instantons) can alter our statements about perturbative dualities, 
in the same vein as in the GLSM approach to mirror symmetry~\cite{Hori:2000kt}. In the latter case, one dualizes the phase of a chiral 
multiplet, hence the system reduces for all purposes to supersymmetric QED with a massless flavor. This theory admits vortex-like instanton 
solutions, generating in the mirror theory a superpotential. 
In contrast, one considers in the present case chiral superfields whose shift-symmetry is gauged, giving after gauge-fixing 
some massive Abelian gauge theory which does not admit instantons.\footnote{Naturally, there are 
instanton solutions associated with the unhatted gauge fields, since the chiral superfields of the base GLSM are minimally coupled to them.} 
This can also be understood by noticing that the target space of the shift superfields has the topology $\mathbb{C}^\star \sim \mathbb{R} \times S^1$. Hence, the previous statements about T-duality are exact; the same conclusion holds for the more generic heterotic dualities that we shall study below. 

%%%%%%%%%%%%%%%%%%%%%%%%%%%%%%%%%%%%%%%%%%%%%%%%%%%%%%%%%%%%%%%%%%%%%%%%%%%%%%%%%%%%%%%%%%%%%%%%%%%%%%%%%%%%%%%%
%%%%%%%%%%%%%%%%%%%TORUS-BUNDLE DUALITIES%%%%%%%%%%%%%%%%%%%%%%%%%%%%%%%%%%%%%%%%%%%%%%%%%%%%%%%%%%%%%%%%%%%%%%%
%%%%%%%%%%%%%%%%%%%%%%%%%%%%%%%%%%%%%%%%%%%%%%%%%%%%%%%%%%%%%%%%%%%%%%%%%%%%%%%%%%%%%%%%%%%%%%%%%%%%%%%%%%%%%%%%

\section{Torus/Bundle dualities}
\label{sec:bund}
In a  $T^2 \hookrightarrow \mathcal{C}_6 \to K3$ compactification, 
one expects a larger duality group, which is a subgroup of $O(2,2+16,\mathbb{Z})$. The extra duality transformations mix the torus fiber  
with Abelian gauge bundles. Implementing directly the relevant duality transformations at the level of the torsional GLSM seems rather difficult, since these dualities exchange classically non-gauge-invariant terms with classically gauge invariant ones suffering 
from quantum gauge anomalies.\footnote{There exists a sort of 'fermionic gauge symmetry',  inherited from the decomposition 
of the $(2,2)$ gauge multiplet into $(0,2)$ multiplets, that can be used to dualize a Fermi multiplet~\cite{Adams:2003zy}, 
but it is not useful in the present context.}

In order to overcome this difficulty, one can bosonize the free charged left-moving worldsheet fermions involved in the duality, as 
suggested in~\cite{Adams:2009av}. We shall explain below how it can be done in the present context of $(0,2)$ GLSMs.

\subsection{(0,2) Bosonization of charged Fermi multiplets}
\label{boso}

For simplicity of the presentation we consider that the non-Abelian gauge bundle $V$ over the $K3$ base lies in the first  $E_8$ factor. Our aim is to find under 
which conditions an $S^1$ fibration can be traded for an Abelian gauge bundle lying in the second $E_8$ factor. 
One starts then with a set of $8$ Fermi multiplets $\Gamma_n$, having vanishing $E$ and $J$ couplings, whose 
charges w.r.t. the $U(1)$ gauge symmetry of the GLSM are labeled by $f_n$. These Fermi multiplets 
are sections of line bundles over $K3$. 

The normalization of the magnetic charges in space-time can be obtained by comparing the 
anomaly cancellation condition in the GLSM with the Bianchi identity in supergravity. In the present GLSM approach, 
an Abelian gauge bundle and the torus fibration contribute generically to the gauge anomaly as 
$f_n f^n+\frac{2U_2}{T_2}|\mathfrak{w}|^2  $ (in units $\alpha'=1$). In supergravity one gets 
the same anomaly equation (see $e.g.$ in~\cite{Carlevaro:2011mn}) by considering 
the pullback of an Abelian bundle over K3 of the form 
$\mathcal{F} = 2\pi k_n \cdot T^n\, \varpi_{[2]}$, where $\varpi_{[2]} \in H^2 (\mathcal{S},\mathbb{Z}) \cap 
H^{1,1} (\mathcal{S})$ is anti-self dual, and $\{T_n, \, n=1,\ldots,8\}$ are  Cartan generators of $SO(16)$, normalized as 
$\kappa (T_n,T_m) = -2 \delta_{nm}$, and with the identification $f_n = 2k_n$.\footnote{One uses 
naturally the same integral basis of anti-self-dual two-forms for defining the torus and gauge bundles, 
corresponding to the intersection matrix $d=(-E_8)\oplus (-E_8) \oplus (-2I_3)$. }

Let us now come back to the bosonization. The {\it on-shell} degrees of freedom of the Fermi multiplets $\Gamma_n$ 
($i.e.$ discarding the auxiliary fields $G_n$ that vanish on-shell), namely left-moving Weyl fermions $\gamma_n$,  
can be embedded into $8$ chiral superfields $B_n$. 
After bosonization, $\gamma_n$ is associated with the left-moving part of $\text{Im} (b_n)$, the 
latter being a compact boson at radius $R_{f} = 1/\sqrt{2}$ in our conventions. Extra degrees of freedom are of 
course necessary in order to get  full chiral multiplets; however, whenever they will be decoupled from the other 
degrees of freedom after duality, it will be possible to safely discard these spectator fields  from the dual theory. 

The original Fermi multiplets generated an anomaly for the unhatted gauge symmetry associated with the 
torsional GLSM, reading  
\begin{equation}
\delta_\Xi \mathcal{L}_{eff} = -\frac{f_n f^n}{8}\int \di \theta^+\, \Xi\,  \Upsilon \,  + h.c. \, .
\end{equation}
Obviously, after bosonization, one should reproduce the same gauge variation, however not as a quantum anomaly 
but as a classical gauge-variance. As for the torus fibration discussed in subsection~\ref{subsec:fibr}, 
the chiral superfields $B_n$ are then coupled {\it both} minimally and axially, with a shift charge $\mathfrak{v}_n$ and an 
axial coupling 
\begin{equation}
\label{AxialBN}
-\frac{ik_n}{4}  \int \di \theta^+ \, \Upsilon\, B_n + h.c. 
\end{equation}
giving the gauge variation
\begin{equation}
\delta_{\Xi} \mathcal{L}_g = \frac{1}{4}   k_n \mathfrak{v}^n  \int \di \theta^+ \, \Upsilon  \Xi + h.c.\, .
\end{equation}
Hence the Bose-Fermi equivalence sets 
\begin{equation}
2k_n \mathfrak{v}^n = - f_n f^n \, .
\end{equation}
On top of this relation, one should keep in mind that, in order to decouple the (non-compact) real part of 
the $b_n$ scalar fields, one should satisfy the relation (see eq.~\ref{chiralcondit}): 
\begin{equation}
\label{radfix}
R_n^{\, 2} \mathfrak{v}_n+ k_n=0 \, ,
\end{equation}
which, given that all the bosons are at the fermionic radius $R_n= 1/\sqrt{2}$, sets (provided that the same should hold for 
{\it any} choice of $\{f_n\}$'s)
\begin{equation}
f_n = -\mathfrak{v}_n = 2 k_n \, ,
\end{equation}
up to an overall sign choice.

At this stage one needs to be careful about the periodicities of the $B_n$'s that are subject to the 
left-moving GSO projection. One has to consider the identifications
\begin{subequations}
\label{gso}
\begin{align}
B_n &\sim B_n + 2i \pi N_n\ , \quad N_n \in \mathbb{Z} \ , \quad  \ 
\sum_{n=1}^{8} N_n \equiv 0 \mod 2  \, , \label{gsoA}\\
B_n &\sim B_n +  i\pi N \ , \quad N \in \mathbb{Z}  \, , \label{gsoB}
\end{align}
\end{subequations}
where the last condition in eq.~(\ref{gsoA}) is imposed by the existence of the Ramond sector, while invariance 
under~(\ref{gsoB}) is a consequence of projecting onto states with even left-moving worldsheet fermion number. 
As a consequence, imposing that the path integral --~in particular the axial coupling~(\ref{AxialBN})~-- 
should be single-valued under the identifications~(\ref{gso}) in any instanton sector, singles out two consistent families of 
line bundles:\footnote{In the $Spin(32)/\mathbb{Z}_2$ theory, these two classes of line bundles would correspond 
respectively to bundles with our without vector structure~\cite{Berkooz:1996iz}.}
\begin{subequations}
\allowdisplaybreaks
\label{kcond}
\begin{align}
\vec{k} &\in \mathbb{Z}^{8} \ , \quad \sum_{n=1}^{8} k_n \equiv 0 \mod 2 \, , \label{kcondA}\\
& \text{or} \notag\\ 
\vec{k} &\in \left(\mathbb{Z}+\frac{1}{2}\right)^{8} \ , \quad \sum_{n=1}^{8} k_n \equiv 0 \mod 2 \, .
\label{kcondB}
\end{align}
\end{subequations}
Consistent gauge bundles $V$  in heterotic are constrained by the vanishing of the second Stiefel-Whitney class~\cite{Witten:1985mj,Freed:1986zx}, namely 
\begin{equation}
\label{chern}
c_1 (V) \in H^{2} (\mathcal{S},2\mathbb{Z}) \, ,
\end{equation}
which gives in the present context (taking into account the normalization of the magnetic charges discussed above) a target-space 
understanding of the condition $\sum_{n=1}^{8} k_n \equiv 0 \mod 2$.

\subsection{Wilson lines}
Bosonizing the left-moving fermions belonging to the Fermi multiplets is also useful in describing the Wilson lines that can be added 
to the $T^2 \hookrightarrow \mathcal{C}_6 \to K3$ solutions, see eq.~(\ref{connecw}). 
After bosonization they correspond in the GLSM to  off-diagonal couplings whose Lagrangian density is of the form
\begin{equation}
\label{superwilson}
L_\textsc{wl} = -\frac{i}{4} \alpha^{a}_n (\Omega_a + \bar{\Omega}_a+2\mathfrak{w}_a \mathcal{A}) 
\partial_- (B^n - \bar{B}^n )\, ,
\end{equation}
where summation over $a=1,2$ and $n=1,\ldots,N$ is implied. Classical gauge invariance requires that 
\begin{equation}
\label{wilsoncoeffs}
\sum_n \alpha^{a}_n \mathfrak{v}_n = 0 \, ,
\end{equation}
whose simplest solution is to have all Fermi multiplets involved being neutral. After refermonization, 
the expression of eq.~(\ref{superwilson}) in components gives couplings like $
\bar{\gamma}_{-}^n \, \alpha^{a}_n\left[ \partial_- (\omega_a -\bar{\omega}_a) + 2 \mathfrak{w}_a A_+ \right]\gamma_-^n$, 
corresponding indeed in space-time to a gauge connection of the form~(\ref{connecw}). The other terms in~(\ref{superwilson}) give 
off-diagonal kinetic terms for the non-compact scalar fields 
$\text{Re} (\omega_a)$ and $\text{Re} (b_n)$ --~which eventually decouple from the theory~-- and for the free worldsheet 
fermions, hence can be rotated away. 

We shall see below that, for special values, the Wilson line moduli $\alpha^{a}_n$ can be traded for 
torus moduli under  T-duality.

\subsection{Gauge-torus duality}
\label{simple}
After these preliminary steps one can move to gauge-torus bundle dualities 
in $T^2 \hookrightarrow \mathcal{C}_6 \to K3$ torsional gauged linear sigma models.

For definiteness we choose, as written above, an $E_8\times E_8$ model such that the non-Abelian component of the gauge bundle 
(that is part of the $(0,2)$ GLSM with $K3$ target space) lies in the first  $E_8$ factor. The second $E_8$ factor is chosen to 
host only an Abelian bundle, corresponding to a set of 8 free Fermi multiplets whose charges are given by a vector $\vec{k}$, 
belonging to one of the two families~(\ref{kcondA},\ref{kcondB}). Our aim is to find how the $T^2$ fibration and the Abelian gauge bundle 
can mix. As we shall see, this duality only makes sense at specific points in the two-torus discrete moduli space. 

To be more specific, we wish to find a perturbative duality exchanging the charges of some bosonized Fermi multiplet $B_{\nu}$
of charge $k_\nu$ and, say,  $\Omega_1$, one of the two chiral multiplets describing the torus fiber over K3, 
that we take neutral, $i.e.$ with a vanishing shift-charge $\mathfrak{w}_1$. In other words, the manifold reduces to a 
circle bundle over $K3$ of topological charge $\mathfrak{w}_2$ times a circle. 

In order to use the technology developed in section~\ref{sec:self}, we can consider an 'auxiliary' complexified two-torus 
made with the chiral superfields $\Omega_1$ and $B_\nu$, whose moduli are set to $\upsilon=i$ and 
$\tau=\nicefrac{i}{2}$, and whose complex topological charge is $\mathfrak{n}=-ik_\nu$. After a $\upsilon\to -1/\upsilon$ duality this 
model is mapped to a dual model of the same type, with the same torus moduli 
(crucially, the dual superfield $\tilde{B}_\nu$ being still at the fermionic radius), and with the dual charge 
\begin{equation}
\tilde{\mathfrak{n}} = - \bar{\upsilon} \mathfrak{n} =  k_\nu\, .
\end{equation}

Compared to the ordinary two-torus case, there is a subtlety  owing to the GSO projection. In deriving the 
duality from a quotient, we had applied the gauge-fixing condition $B_\nu=0$, see around eq.~(\ref{Idual}). 
However, since this superfield is subject to the identifications~(\ref{gso}), there exists corresponding 
residual large gauge transformations that are not fixed in this gauge. It boils down to assigning to $\tilde{\Omega}_1$, the 
superfield dual to $\Omega_1$, a shift-charge under the GSO projection. This seems consistent with the analysis of~\cite{Adams:2009zg}, 
where a detailed analysis of the torsional GLSMs under consideration was done. It was shown there that the shift-superfields of the 
torus fibration must carry a non-zero shift-charge under the global $U(1)_L$ symmetry which is used in defining the left 
GSO projections. At the same time, the dual superfield $\tilde{B}_\nu$ is expected to inherit the GSO charge assignment of $\Omega_1$.  
It would be useful to study this point in more depth in order to understand its target-space interpretation.\footnote{It is worthwhile to 
remind that the condition~(\ref{chern}), which plays an important role in this discussion, 
is necessary as the bundle should admit spinors, which belong to the massive sector of heterotic strings, hence are 'invisible' to supergravity.}

After refermionizing $\tilde{B}_\nu$ and throwing away the decoupled extra degrees of freedom belonging to this  
chiral multiplet, one gets the pullback of an Abelian gauge bundle on K3 whose embedding $\vec{k}$ in the second $E_8$ is such 
that its $\nu$-th component vanishes, and a two-torus principal bundle of topological charge 
$\mathfrak{w} = k_\nu + T \mathfrak{w}_2$. Hence T-duality relates two solutions with different topologies.  

In retrospect, looking back at the analysis of line bundles done in subsection~\ref{boso}, the dual model cannot be interpreted 
as describing a standard geometric $T^2 \hookrightarrow \mathcal{C}_6 \to K3$ compactification for every value of $k_\nu$. 
As $k_\nu$ needs to be integer in order to get a proper geometrical interpretation, only bundles of the 
type~(\ref{kcondA}) can be chosen as starting points. Furthermore, as we wish that the condition 
$c_1 (V) \in H^{2} (\mathcal{S},2\mathbb{Z})$ still holds after the duality, one needs to consider only the cases 
$k_\nu \in 2 \mathbb{Z}$. 

The worldsheet theories should still make sense otherwise, since the modified 
GSO projection that is obtained after duality, see the discussion above, eventually gives consistent models; 
however they won't have a straightforward geometric interpretation.

\subsubsection*{Non-orthogonal tori and Wilson lines}
As $U_1$ and $T_1$, the real part of the K\"ahler and complex structures of the fibered two-torus, 
are generically non-zero, there are extra non-diagonal terms appearing in the dual theory, that we wish to identify as Wilson lines 
similar to~(\ref{superwilson}). We need to avoid the appearance of terms like $(B_\nu + \bar{B}_\nu) 
\partial_- (\Omega_2 - \bar{\Omega}_2)$, since the real part of $B_\nu$ should eventually 
decouple  from the torsional GLSM. This is granted whenever $U_1 = 2  T_1$, leading, as $R_1^{\, 2} = U_2/T_2 = 2$, to
\begin{equation}
\label{bundledualcondi}
U = 2  T\, ,
\end{equation}
which is a condition for the perturbative duality exchanging the torus fibration and the Abelian gauge bundle 
to exist. Whenever the consistency condition~(\ref{bundledualcondi}) is satisfied, the dual Lagrangian 
density contains a term 
\begin{equation}
\label{dualtermssimplwilson}
-\frac{i}{4}T_1   (\Omega_2 + \bar \Omega_2 +2 \mathfrak{w}_2 \mathcal{A}) 
\, \partial_- (B_\nu - \bar{B}_\nu), 
\end{equation}
hence the gauge bundle of the dual theory embedded in the second $E_8$ is the Whitney sum of the pullback of the Abelian 
gauge bundle over the K3 base discussed above and of a line bundle over the total space $\mathcal{C}_6$, 
given by a Wilson line around the one-cycle of the fiber corresponding to $\Im(\omega_2)$ with 
$\alpha^2_\nu = T_1  \in \mathbb{Q}$. 
		
It is possible to get rid of the remaining $S^1$ fibration over the $K3$ base as well, by further dualizing the model. 
Other, more complicated examples are naturally possible by considering dualities involving also free Fermi multiplets from 
the first $E_8$ factor, or $Spin(32)/\mathbb{Z}_2$-based constructions.

\subsubsection*{Wilson line moduli and duality}

Perturbative dualities map $T^2 \hookrightarrow \mathcal{C}_6 \to K3$ compactifications with some HYM gauge bundle to $K3\times T^2$ compactifications with 
an extra Abelian gauge bundle. In the former model the torus moduli $(U,T)$ are generically quantized, 
belonging to the same number field $\mathbb{Q}(\sqrt{D})$, while in the latter model they are free to take any value. 
The upshot is that the torus moduli are traded for gauge ($i.e.$ Wilson lines) moduli.

To illustrate this point, let us consider the simplest example of gauge/torus duality. The starting point is an $S^1 \times K3$ compactification, 
with a line bundle for which only the component $k_\nu$ of the charge vector is non-vanishing. In this model, 
while the radius of the circle is obviously a free moduli, there are only seven independent Wilson lines 
moduli from the second $E_8$ factor, see the constraint~(\ref{wilsoncoeffs}). The dual model is an 
$S^1 \hookrightarrow \mathcal{C}_5 \to K3$ compactification, without gauge bundle in the second $E_8$ factor. In this case there are 
no constraints on the Wilson lines, while the radius of the circle fiber is quantized as $R^2 \in \mathbb{Q}$. Therefore, the 
dimension of the connected component of the moduli space containing the circle-gauge bundle duality point is the same on both sides, 
as it should since these are two descriptions of the same physics.

%%%%%%%%%%%%%%%%%%%%%%%%%%%%%%%%%%%%%%%%%%%%%%%%%%%%%%%%%%%%%%%%%%%%%%%%%%%%%%%%%%%%%%%%%%%%%%%%%%%%%%%%%%%%%%%%%%%%%%%%%%%%%%%%%%%%%%%%%%%%%%%%%%%%%%%%%%%%%

\section{Conclusion}
\label{sec:conc}

In this work we have studied perturbative dualities in torsional principal torus bundles compactifications of the heterotic string. 
Using a gauged linear sigma-model approach allowed us to prove that these dualities, that one can infer from the Buscher rules whose 
validity in this context is not granted, are exact symmetries of the theory. 

The first type of duality maps the torus fiber onto itself and gives a model lying in the same class as the original one, provided 
that the Kaluza-Klein charges defining the fibration are transformed accordingly. A very interesting outcome is that both moduli of 
the two-torus, the complex and K\"ahler structure, should be quantized in a precise way, otherwise duality transformations acting on 
the K\"ahler structure wouldn't make sense for every choice of topological charge. 
While these quantization conditions are rather familiar in the type IIB 
orientifolds duals, it is interesting to see them arising in a completely different way in the heterotic sigma-model. 

The second type of duality, mixing the torus bundle with Abelian gauge bundles, is particularly fascinating since it relates 
solutions with different topologies. At the gauged linear sigma-model level it maps the torsional GLSMs introduced recently 
to more familiar ordinary $(0,2)$ GLSMs whose gauge bundle has an Abelian component;  
hence there isn't any sharp distinction between these two classes of models. It would be interesting to understand 
better to which extent the interplay of the duality with the generalized left-moving GSO projection that we mentioned in the text 
could blur the geometrical interpretation of this duality. 

Compared to the ordinary $K3\times T^2$ compactifications, the $T^2 \hookrightarrow \mathcal{C}_6 \to K3$ solutions 
have two distinguished features, besides the torus fibration itself. As follows from supersymmetry, the CY base is warped, and 
there is a non-zero three-form flux. One can wonder what happens to these features under a torus-gauge duality. 
With a non-compact Eguchi-Hanson base, for which both a worldsheet CFT and a proper supergravity limit 
are available~\cite{Carlevaro:2008qf}, the answer can be found. The warp factor stays the same, while the NSNS three-form looses 
its components with a leg along the torus but doesn't vanish. In the compact case, considering $K3\times T^2$ 
with a departure from the standard embedding, satisfying the tadpole condition~(\ref{tadpole}), 
a Ricci-flat metric on $K3$ is a solution of the supersymmetry equations at tree level in the $\alpha'$ expansion, while 
departure from the CY condition as well as torsion are expected to arise at higher order in $\alpha'$. 
The fibered solutions  are built upon a very different ansatz, including torsion and warping from the start, but, as the 
tadpole condition~(\ref{tadpole}) forbids the two-torus volume from being large in string units, the distinction between 
both type of solutions in supergravity, as far as the base is concerned, is not sharply defined. 

This duality has also interesting consequences on the effective four-dimensional actions deriving from these 
compactifications. For instance, the one-loop threshold corrections to the gauge couplings is well-known for 
$K3\times T^2$ compactifications with or without Wilson lines, being a derived product of the elliptic genus 
(see $e.g.$~\cite{Camara:2008zk} and references therein). Provided that the torus and Wilson line moduli are mapped through 
the duality according to the rules derived from our analysis, the same quantity gives also the one-loop corrections 
for compactifications on torsional $T^2 \hookrightarrow \mathcal{C}_6 \to K3$ manifolds.

Finally, an ambitious goal would be to find similar duality relations between torsional and torsion-free heterotic solutions 
with $\mathcal{N}=1$ supersymmetry in four dimensions, more specifically between $SU(3)$-holonomy and $SU(3)$-structure compactifications. This requires 
a good handle on $(0,2)$ mirror symmetry since, in this case, worldsheet instantons are expected to play an important role. We leave 
this analysis for future work.

\section*{Acknowledgements}
I would like to thank Luca Carlevaro, Amit Giveon, Stefan Groot-Nibbelink, Nick Halmagyi, Josh Lapan and Jan Troost for 
enlightening discussions. This work was supported by French state funds managed by the ANR within the Investissements d'Avenir 
programme under reference ANR-11-IDEX-0004-02.

%%%%%%%%%%%%%%%%%%%%%%%%%%%%%%%%%%%%%%%%%%%%%%%%%%%%%%%%%%APPENDIX%%%%%%%%%%%%%%%%%%%%%%%%%%%%%%%%%%%%%%%%%%%%%%%%%%%%%%%%%%%%%%%%%%%

\appendix

\section{Two-dimensional $(0,2)$ superspace}

Minkowskian $(0,2)$ superspace is spanned by the coordinates $(x^+,x^-, \theta^+,\bar \theta^+)$ with 
$\bar{\theta}^+ = (\theta^+)^\dag$. The associated Berezin integral reads (with $\di^2 \theta = \di \bar{\theta}^+ \di \theta^+$):
\begin{equation}
\int \di^2 \theta\, \theta^+ \bar\theta^+ = 1\, .
\end{equation}
We define the super-space derivatives and super-charges as follows:
\begin{subequations}
\begin{align}
\mathcal{Q}_+ &= \partial_{\theta^+} +i \bar\theta^+ \partial_+ \quad , \qquad \bar{\mathcal{Q}}_+ = -\partial_{\bar\theta^+} -i \theta^+  \partial_+ \, , \\
D_+ &= \partial_{\theta^+} -i \bar\theta^+ \partial_+ \quad , \qquad \bar{D}_+ = -\partial_{\bar\theta^+} + i\theta^+ \partial_+  \, .
\end{align}
\end{subequations}
The non-trivial anti-commutators are then
\begin{equation}
\{ \bar{D}_+,D_+\} = 2 i \partial_+  \quad , \qquad \{ \bar{\mathcal{Q}}_+,\mathcal{Q}_+\} = -2i \partial_+
\end{equation}

\subsection{Superfields}
\label{app:fields}
We give the component expansion of the superfields that are needed in this work. 
\paragraph{Chiral superfields} are defined by the constraint
\begin{equation}
\bar{D}_+ \Phi = 0
\end{equation}
Hence the chiral superfield reads in components
\begin{equation}
\Phi = \phi + \sqrt{2}\theta^+ \lambda_+ -i \theta^+ \bar\theta^+  \partial_+ \phi
\end{equation}

\paragraph{Fermi superfields} have as a bottom component a left-moving fermion. They satisfy generically the constraint
\begin{equation}
\label{fermiconstr}
\bar{D}_+ \Gamma = \sqrt{2} E (\Phi) \, ,
\end{equation}
where $E$ is an holomorphic function. Hence have the component expansion
\begin{equation}
\Gamma =\gamma_- + \sqrt{2} \theta^+ G - \sqrt{2} \bar\theta^+ E(\Phi) -i \theta^+ \bar\theta^+ \partial_+ \gamma_-\, ,
\end{equation}
where $G$ is an auxiliary field.

\paragraph{Gauge multiplets} are actually described by a pair of $(0,2)$ superfields, namely $\mathcal{A}$ and 
$\mathcal{V} $. 
Super-gauge transformations act as
\begin{equation}
\mathcal{A} \to \mathcal{A} +\frac{i}{2}(\bar \Xi - \Xi) \quad , \qquad 
\mathcal{V}  \to \mathcal{V}  -\frac{1}{2} \partial_- (\Xi + \bar \Xi) 
\end{equation}
where $\Xi$ is a chiral superfield.

In the Wess-Zumino gauge things get simpler, even though one should be careful while dealing with 
classically non gauge-invariant actions. The residual gauge symmetry is given in this gauge by
\begin{equation}
\Xi = \rho -i \theta^+ \bar\theta^+ \partial_+ \rho
\end{equation}
with real $\rho$, while the component expansion of $\mathcal{A}$ and $\mathcal{V}$ read
\begin{subequations}
\begin{align}
\mathcal{A} &= \theta^+ \bar{\theta}^+ A_+ \\
\mathcal{V}  & = A_- -2 i \theta^+ \bar{\mu}_- -2i\bar{\theta}^+ \mu_- + 2 \theta^+ \bar{\theta}^+ D
\end{align}
\end{subequations}
where $D$ is an auxiliary field. Accordingly 
the components $A_{\pm}=A_0 \pm A_1$ of the gauge field are shifted under the residual gauge transformations as
\begin{equation}
A_\pm \stackrel{\rho}{\longrightarrow} A_\pm - \partial_\pm \rho  \\
\end{equation}

In order to couple the gauge superfields to the chiral and Fermi superfields we first define ordinary covariant derivatives as
\begin{equation}
\nabla_{\pm} = \partial_{\pm} +i Q A_{\pm}\, .
\end{equation}
Then we construct the gauge-covariant superderivatives as follows:
\begin{subequations}
\begin{align}
\mathfrak{D}_+ &= (\partial_{\theta^+} -i \bar\theta^+ \nabla_+) = D_+ +Q \bar \theta^+  A_+ \\
\bar{\mathfrak{D}}_+ &= (-\partial_{\bar{\theta}^+} +i \theta^+ \nabla_+) = \bar{D}_+ -Q \theta^+  A_+ 
\end{align}
\end{subequations}
They satisfy the algebra
\begin{equation}
\{ \mathfrak{D}_+, \bar{\mathfrak{D}}_+ \} = 2 i \nabla_+\, .
\end{equation}
We also need the superspace version of the gauge-covariant derivative, since our theory is chiral. It is defined as
\begin{equation}
\mathcal{D}_- = \partial_- + i Q \mathcal{V}  \, 
\end{equation}
whose lowest component is $\nabla_-$. 

\paragraph{Charged matter multiplets}
A chiral multiplet of charge $Q$ needs to satisfy the gauge-covariant constraint:
\begin{equation}
\bar{\mathfrak{D}}_+ \Phi = 0
\end{equation}
which is solved by
\begin{equation}
\label{chiralcov}
\Phi = \phi + \sqrt{2}\theta^+ \lambda_+ -i \theta^+ \bar\theta^+  \nabla_+ \phi \, .
\end{equation}
In other words, since 
\begin{equation}
\bar{\mathfrak{D}}_+ = e^{Q \mathcal{A}} \bar{D}_+ e^{-Q \mathcal{A}} 
\end{equation}
We have that 
\begin{equation}
\Phi = e^{Q \mathcal{A}} \Phi_0
\end{equation}
where $\Phi_0$ is a superfield obeying the standard chirality constraint $\bar{D}_+ \Phi_0=0$.

Similarly, a charged Fermi superfield of charge $q$ can be obtained as $\Gamma = e^{q \mathcal{A}} \Gamma_0$ where $\Gamma_0$ 
satisfies $\bar{D}_+ \Gamma_0 = \sqrt{2} E$. Hence the superfield $\Gamma$ has the component expansion:
\begin{equation}
\Gamma =\gamma_- + \sqrt{2} \theta^+ G - \sqrt{2} \bar\theta^+ E(\Phi) -i \theta^+ \bar\theta^+ \nabla_+ \gamma_-\, ,
\end{equation}
where as before $E$ is an holomorphic function in the chiral superfields.

\subsubsection*{Shift chiral superfield}
We  define a {\it shift chiral superfield} $\Omega$, which is a chiral superfield which transforms under gauge transformations as
\begin{equation}
\label{gaugeomega}
\Omega  \stackrel{\Xi}{\longrightarrow} \Omega + i \mathfrak{w} \Xi \, .\\
\end{equation}
The components expansion of this superfield is simply 
\begin{equation}
\Omega = \omega + \sqrt{2}\theta^+ \chi_+ -i \theta^+ \bar\theta^+  \partial_+ \omega \, ,
\end{equation} 
such that it satisfies the standard chirality constraint. Notice that only the imaginary part of $\Omega$ is 
shifted by the gauge transformation~(\ref{gaugeomega}).

%%%%%%%%%%%%%%%%%%%%%%%%%%%%%%%%%%%%%%%%%%%%%%%%%%%%%%%%%%%%%%%%%%%%%%%%%%%%%%
\subsection{(0,2) Lagrangians}
\label{app:lag}

Let us start with the kinetic term for a chiral field $\Phi$ of charge $Q$, whose components expansion 
is given by~(\ref{chiralcov}). It is given by 
\begin{subequations}
\begin{align}
\mathcal{L}_s &= -\frac{i}{2} \int \di^2 \theta^+ \,  \bar{\Phi} \mathcal{D}_- \Phi\\
&= \frac{1}{2} \left( \overline{\nabla_+ \phi} \nabla_- \phi + \nabla_- \bar{\phi}  \nabla_+ \phi \right)
+ i \bar{\lambda}_+ \nabla_- \lambda_+ 
+ i \sqrt{2} Q \left( \lambda_+ \mu_- \bar{\phi} + h.c. \right)
+ Q \phi \bar{\phi} D\, \, .
\end{align}
\end{subequations}

Let us now move on to the case of a Fermi superfield of charge $q$. One has the following component expansion:
\begin{subequations}
\begin{align}
\mathcal{L}_f &= -\frac12 \int \di^2 \theta^+ \,  \bar{\Gamma} \Gamma \\
&=i\bar{\gamma}_- \nabla_+ \gamma_- 
+ |G|^2 - |E(\phi)|^2 -\left(E'(\phi) \bar{\gamma}_- \lambda_+ + h.c. \right)
\end{align}
\end{subequations}

The gauge kinetic term is written in terms of the field strength superfield, which is a chiral superfield
\begin{equation}
\label{upsidef}
\Upsilon = \bar{D}_+ (\partial_- \mathcal{A}+i \mathcal{V} ) = -2 \left(\mu_- -i \theta^+ (D-i F_{01} )
-i\theta^+ \bar{\theta}^+ \partial_+ \mu_- \right)
\end{equation}
with $2F_{01}=\partial_- A_+ - \partial_+ A_-$. Including a possible constant FI term, of parameter 
$t=ir+ \frac{\theta}{2\pi}$, one has 
\begin{subequations}
\begin{align}
\mathcal{L}_g &= -\frac{1}{8e^2} \int \di^2 \theta^+ \, \bar{\Upsilon} \Upsilon + \frac{t}{4}  \int \di \theta^+ \, 
\Upsilon + h.c. \\
&= \frac{1}{2e^2} \left( 2i \bar{\mu}_- \partial_+ \mu_-  + D^2+F_{01}^2 \right)
-r D + \tfrac{\theta}{2\pi} F_{01} 
\end{align}
\end{subequations}
Due to the presence of chiral fermions there is generically a gauge anomaly. It is given as 
\begin{equation}
\delta_\Xi W = \frac{\mathfrak{A}}{16 \pi} \int \di^2 x \, \int \di \theta^+\, \Xi\,  \Upsilon \,  + h.c. \, ,
\label{anomaly}
\end{equation}
with $\mathfrak{A}=Q_i Q^i - q_n q^n$.

Finally, a gauge-invariant kinetic term for the shift chiral superfield can be constructed in superspace as follows
\begin{subequations}
\begin{align}
\mathcal{L}_s  =& -\frac{i}{4} \int \di^2 \theta \left( \Omega + \bar \Omega+2 m \mathcal{A} \right)
\left(\partial_- (\Omega-\bar \Omega)+ 2i\mathfrak{w} \mathcal{V}  \right)\\
=&\frac{1}{4} \partial_+(\omega+\bar{\omega})  \partial_- (\omega+ \bar{\omega})-
\frac{1}{4} \left( \partial_+ (\omega-\bar \omega) + 2i m A_+ \right)\left( \partial_- (\omega-\bar \omega) + 2i m A_- \right)\notag \\
& + \frac{i}{2} \left(\chi_+ \partial_- \bar{\chi}_+ + \bar{\chi}_+ \partial_- \chi_+\right) + i \sqrt{2} m (\chi_+ \mu_- 
+ \bar{\chi}_+ \bar{\mu}_- )+ m D(\omega+\bar \omega)\, .
\end{align}
\end{subequations}
It is possible to add a FI-like term (axial coupling):
\begin{subequations}
\begin{align}
\mathcal{L}_{g}  &= -\frac{ih}{4}  \int \di \theta^+ \,  \Omega \Upsilon + h.c. \\
&= h D \, \text{Re}\, (\omega) + h F_{01}\, \text{Im}\, (\omega) 
- \frac{ih}{\sqrt{2}} \left(\mu_- \chi_+-\bar{\mu}_- \bar{\chi}_+\right)
\end{align}
\end{subequations}
Whenever the shift superfield is charged, this coupling is classically not gauge invariant. Indeed
\begin{equation}
\delta_{\Xi} \mathcal{L}_g = \frac{h}{4}  \mathfrak{w}  \int \di \theta^+ \,  \Xi \Upsilon + h.c.
\end{equation}
This terms combines nicely with the one-loop anomaly, eq~(\ref{anomaly}) in order to get a gauge-invariant theory. 
The axial coupling can also be rewritten, in order to facilitate the duality transformations, as an integral over the whole superspace, 
using eq.~(\ref{upsidef}): 
\begin{equation}
\label{fullsuper}
\mathcal{L}_g =  \frac{h}{4}\int \di^2 \theta^+ \, \left[\mathcal{V}  (\Omega+\bar\Omega) 
+ i \mathcal{A} \partial_- (\Omega-\bar\Omega)
\right]+ t.d. \,.
\end{equation}
Notice that one could also consider an ordinary charged chiral superfield $\Phi$, with a logarithmic FI-like term, 
leading to similar effects:
\begin{equation}
\mathcal{L}_{g} '  = -\frac{ih}{4}  \int \di \theta^+ \, \Upsilon \log \tilde \Phi + h.c. 
\end{equation}

\subsubsection*{Superpotential terms}
The last term in the GLSM is the superpotential term, given by a set of holomorphic functions $J$ of the 
chiral superfields. It has to satisfy the constraint $E^a J_a=0$ 
(where $a$ runs over the Fermi superfields). It reads
\begin{subequations}
\begin{align}
\mathcal{L}_p  &= -\frac{\mu}{2}  \int \di \theta^+ \, 
\Gamma J + h.c. \\
&= \frac{\mu}{\sqrt{2}} (G J + h.c.) + \frac{\mu}{\sqrt{2}}  \gamma_- \left( \lambda_+\partial_\phi J  + \chi_+\partial_\omega J  \right) 
+ h.c. 
\end{align}
\end{subequations}
After solving for the auxiliary fields of the full theory one gets the scalar potential
\begin{equation}
U(\phi,\omega) = \frac{e^2}{8}\left( Q |\phi|^2 -r \right)^2 + \frac{\mu}{2} |J|^2  + |E|^2\, ,
\end{equation}
which defines the vacua of the theory $Q|\phi|^2 =r$, $J=0$ and $E=0$ modulo gauge transformations.

\section{Two-tori}
\label{apptor}
The moduli space of $T^2$ compactifications is spanned 
by the vacuum expectation values of two complex fields
\begin{equation}
T = \frac{G_{1 2}+i\sqrt{\det G}}{G_{1 1}} \quad , \qquad U = B_{1 2} + i \sqrt{\det G}
\end{equation}
where the coordinates $(x^1,x^2)$ are $2\pi$-periodic. In terms of these parameters the metric is rewritten conveniently as 
\begin{equation}
\di s^2 = \frac{U_2}{T_2} | \di x^1 + T \di x^2 |^2
\end{equation}
It is natural to gather the metric and the B-field in a single matrix
\begin{equation}
E = G+B = \frac{U_2}{T_2} \left( \begin{array}{cc} 1 & T_1 + T_2 U_1 / U_2\\
T_1 - T_2 U_1 / U_2 & T \bar T  \end{array} \right)
\end{equation}

String compactifications on a two-torus have a common $O(2,2,\mathbb{Z})$ perturbative duality group which decomposes as
\begin{equation}
PSL(2,\mathbb{Z}) \times PSL(2,\mathbb{Z}) \times \zi_2 \times \zi_2
\end{equation}
The first $PSL(2,\mathbb{Z})$ acts on $T$ as 
\begin{equation}
T \to \frac{aT+b}{cT+d} \quad , \qquad ad-bc=1 \quad , \qquad a,b,c,d \in \zi
\end{equation} 
and leaves $U$ invariant. It corresponds to the familiar modular group of the two-torus (which 
does not contain any 'stringy' symmetry).

The second $PSL(2,\mathbb{Z})$ acts on $U$ in a similar way, leaving $T$ invariant; it contains e.g. integer
shifts of the $B$-field. It generalizes the usual T-duality of $S^1$ compactifications.

The extra $\zi_2$ symmetries are the spacetime parity $(U,T) \to (- \bar U,-\bar T)$ 
and  T-duality along $x^1$, which exchanges the generalized K\"ahler modulus and the complex structure
\begin{equation}
(T;U) \stackrel{T_x}{\longrightarrow} (U;T)\, ,
\end{equation}
hence is nothing but mirror symmetry in one complex dimension.

\bibliography{bibtdual}

\end{document}